\documentclass[12pt,letterpaper]{article}

\usepackage{amsmath}
\usepackage[psamsfonts]{amssymb}
\usepackage{amsthm}
\usepackage{fullpage}
\usepackage{cite}
\usepackage{indentfirst}
\usepackage{epsf}

\usepackage{multirow}

\newcommand{\R}{{\mathbb R}}

\newcommand{\Z}{{\mathbb Z}}

\newcommand{\half}{\ensuremath{\frac{1}{2}}}
\newcommand{\qtr}{\ensuremath{\frac{1}{4}}}

\newcommand{\PP}{{\mathbb P}}

\newcommand{\CC}{{\mathcal C}}
\newcommand{\FF}{{\mathbb F}}

\newcommand{\IP}[1]{\langle#1\rangle}
\newcommand{\iso}{\simeq}

\newcommand{\pl}{\widetilde{l_p}}

\newcommand{\B}{\mathbb{B}}

\newcommand{\M}{\mathcal{M}}
\newcommand{\h}{\mathfrak{h}}
\newcommand{\mhk}{\widehat{\M_k}}
\newcommand{\mk}{\M_k}
\newcommand{\degC}{d_{{\cal C}}}

\newcommand{\onefigure}[2]{\begin{figure}[htbp]
         \caption{#2\label{#1}(#1)}
         \end{figure}}

\renewcommand{\onefigure}[2]{\begin{figure}[htbp]
         \begin{center}\leavevmode\epsfbox{#1.eps}\end{center}
         \caption{#2\label{#1}}
         \end{figure}}
\newcommand{\comment}[1]{}

\newcommand{\figref}[1]{Fig.~\protect\ref{#1}}

\def\bbbz{{\sf Z\!\!\!Z}}
\def\sl2z{SL(2,\bbbz)}
\newcommand{\be}{\begin{equation}}
\newcommand{\ee}{\end{equation}}
\newcommand{\bea}{\begin{eqnarray}}
\newcommand{\eea}{\end{eqnarray}}

\def\half{\frac{1}{2}}

\def\bbbz{{\sf Z\!\!\!Z}}
\def\sl2z{SL(2,\bbbz)}

\def\z0{{\bf z_0}}

\begin{document}

\bibliographystyle{utphys}

\setcounter{page}{1}
\pagestyle{plain}

\begin{titlepage}

\begin{center}
\today
\hfill HUTP-01/A053\\
\hfill UTTG-16-01\\
\hfill                  hep-th/0111068

\vskip 3 cm
{\large \bf A Mysterious Duality}
\vskip 2 cm
{Amer Iqbal$^{~1}$, Andrew Neitzke$^{~2}$ and Cumrun Vafa$^{~3}$}\\
\vskip 0.5cm
{$^{1}$Theory Group, Department of Physics\\
University of Texas at Austin,\\
Austin, TX 78712, U.S.A.\\}
\vskip 0.5cm
$^{2}${Department of Mathematics \\
Harvard University\\
Cambridge, MA 02138, U.S.A.\\}
\vskip 0.5cm
$^{3}${ Jefferson Physical Laboratory,
\\Harvard University\\
Cambridge, MA 02138, U.S.A.\\}

\end{center}

\vskip 2 cm
\begin{abstract}
We establish a correspondence between toroidal
compactifications of M-theory
and del Pezzo surfaces.
M-theory on $T^k$ corresponds to
$\PP^2$ blown up at $k$ generic points; Type IIB corresponds
to $\PP^1\times \PP^1$.
The moduli of compactifications of M-theory
on rectangular tori are mapped to K\"ahler moduli of del Pezzo
surfaces.  The U-duality group of M-theory corresponds to
a group of classical symmetries of the del Pezzo represented by global
diffeomorphisms.
The $\half$-BPS brane charges of M-theory correspond to
spheres in the del Pezzo, and their tension to the exponentiated
volume of the corresponding spheres.  The electric/magnetic
pairing of branes is determined by the condition that the union of the
corresponding spheres represent the anticanonical class of the del Pezzo.
The condition that a pair of $\half$-BPS states
form a bound state is mapped to a condition
on the intersection of the corresponding spheres.  We present some
speculations about the meaning of this duality.
\end{abstract}

\end{titlepage}

\section{Introduction}

The discovery of duality symmetries in string theory has led
to spectacular progress in our understanding of non-perturbative
aspects of the theory.  However, we still do not have a deep
understanding of the meaning of these symmetries.  Indeed, we find ourselves
in the strange situation that the quantum corrected physics
has {\it more} symmetries than the classical strings had any right
to expect!

A clear appreciation of symmetry principles is a sacred principle of physics.
Given any physical system, we should formulate the theory in
a way that makes all of the symmetries manifest.  But despite
many years of work on a non-perturbative definition of
string theory, we are no closer to making duality symmetries
manifest than when the dualities were first discovered!

We would like to have a formulation of string theory in which all of
the duality symmetries are classically visible.
The aim of this paper is to develop a mysterious duality which points
to the existence of such a formulation.

It was noted in \cite{vafa-talk} that there is a classical geometric
system which shares all the U-duality symmetries of M-theory
compactified on rectangular tori.  The relevant geometric
objects are del Pezzo surfaces, which are complex 2-dimensional
K\"ahler manifolds with $c_1>0$.
In this correspondence, M-theory
in 11 dimensions is mapped to $\PP^2$, and subsequent compactifications
of M-theory on rectangular tori $T^k$ are mapped
to $\PP^2$ blown up at $k$ generic points (i.e. a manifold in which $k$ 
generic points on $\PP^2$ are replaced with $\PP^1$'s),
which is a del Pezzo surface denoted $\B_k$.
Type IIB is mapped to $\PP^1\times \PP^1$,
which can be viewed as a blow down of $\B_2$.  The U-duality group
of M-theory on $T^k$, which for rectangular compactifications with no $C$-field
vevs is given by the Weyl group of $E_k$
(with a suitable definition for small $k$), is mapped to a subgroup of
the global diffeomorphisms
of the del Pezzo.  For example, the $S$-duality of type IIB in 10 dimensions
is realized as the exchange of the two $\PP^1$'s in $\PP^1\times \PP^1$.

We will show in this paper that one can take this idea quite far,
constructing a precise dictionary relating the two sides.  In particular, the
$\half$-BPS $p$-branes of M-theory will be mapped to rational curves
on the del Pezzo.  Furthermore, we find
a map which relates the moduli of the (extended) K\"ahler metric
on $\B_k$, which is determined by $k+1$ real parameters (one
for the overall size of the original $\PP^2$ 
and $k$ for the volumes of the blown-up $\PP^1$'s)
with the $k+1$ moduli of M-theory on $T^k$
($k$ moduli for the compactification radii and one for
the Planck scale).
The moduli space of K\"ahler classes also carries
an action of a group of global diffeomorphisms of the del Pezzo surface,
which exchange the various embedded 2-spheres; as noted above,
this group is isomorphic to the Weyl group of $E_k$.  Moreover,
this symmetry action is compatible
with the identification of the moduli on the two sides.
The map relates the tension of a $\half$-BPS
state with the exponential of the volume of the corresponding
$\PP^1$.  In addition, electric/magnetic duality and
the condition for a pair of branes to form a bound state
admit a nice geometric interpretation on the del Pezzo side.

Some aspects of the dictionary we construct
are listed in the table below:

\vglue 0.5cm

\noindent \begin{tabular} {||c|c||}\hline
del Pezzo $\B_k$ & M-theory on $T^k$ \\ \hline\hline
Element of $H^2(\B_k, \R)$ & Point in moduli space of M-theory on $T^k$ \\ \hline
Global diffeomorphisms & \multirow{2}{*}{U-duality group} \\
 preserving the canonical class $K$ & \\ \hline
2-sphere $\CC$ with volume $V_{\cal C}$ and  & \multirow{2}{*}{$\half$-BPS $p$-brane state with tension $2 \pi \exp V_C$} \\
degree $p+1$ & \\ \hline Volume $V_K$ of canonical
class & Compactified Planck length, $\pl^{9-k} = \exp V_K$ \\
\hline Volume $V_H$ of hyperplane class & 11-dimensional Planck
length, $l_p^{-3} = \exp V_H$ \\ \hline Volume $V_E$ of exceptional
curve & Radius $2\pi R = \exp -V_E$ \\
\hline $H$, line in $\PP^2$ &
M2-brane \\ \hline 2$H$, conic in $\PP^2$ & M5-brane \\ \hline
2-spheres $\CC_1$, $\CC_2$ with $\CC_1 + \CC_2 = -K$ & Electric-magnetic
dual objects \\ \hline
\end{tabular}

\vglue 0.5cm

Here $H$ denotes the class of a hyperplane (a complex line) in $\PP^2$, and
$K$ denotes the ``canonical class,'' a 2-cycle class which is
dual to the negative of the first Chern class of the del Pezzo.

The organization of this paper is as follows:  In Section 2 we
review some relevant geometric aspects of del Pezzo surfaces.
In Section 3 we present the map between the two sides.  In Section
4 we discuss some speculations as to the meaning of this duality
and raise some natural questions.

\section{Del Pezzo surfaces}

\subsection{Basic properties} \label{homology}

The term ``del Pezzo surface'' refers to any manifold of
complex dimension $2$ such that the first Chern class is positive
\cite{MR87d:11037}. These
surfaces admit metrics of positive scalar curvature.
They may be classified as follows:  either take
$\PP^2$ and blow up $k \le 8$ generic points, or take $\PP^1 \times \PP^1$
and blow up $k \le 7$ generic points.  We call the resulting del Pezzos
$\B_k$ and $\FF^k$ respectively.  In fact $\B_{k+1} \iso \FF^k$
for $k \ge 1$, so it is enough to consider only $\B_k$ and $\PP^1 \times \PP^1$.

Now let us describe the homology of del Pezzo surfaces, beginning with $\PP^2$.  
Since $\PP^2$ is simply connected the only interesting homology will be in
dimension 2; in fact $H_2(\PP^2, \Z)$ is simply $\Z$, generated by the
class of a line, which we write $H$.  When we blow up a point we replace it
with a $\PP^1$ (a 2-sphere), which gives a new generator in $H_2(\B_k, \Z)$ for
each of the $k$ points we blow up.  Thus
$\mbox{dim}\ H_{2}(\mathbb{B}_{k}, \R)=k+1$; a natural basis to choose is
$\{H,E_{1},\dots, E_{k}\}$, where the $E_{a}$ are the ``exceptional curves''
obtained by blow-ups and $H$ represents the pullback of the generator of
$H_{2}(\PP^2, \Z)$ under the projection \bea
\pi\,:\,\mathbb{B}_{k}\to \mathbb{B}_{0}=\mathbb{P}^{2}  \eea
which simply collapses each exceptional curve
$E_{a}$ to the corresponding point $p_{a}\in \mathbb{P}^{2}$ which was blown up. The
intersection numbers are given
by \cite{MR57:3116} \bea ^{}H\cdot H=1\,,\,\,^{}H\cdot
E_{a}=0\,,\,\, ^{}E_{a}\cdot
E_{b}=-\delta_{ab}\,,\,a,b=1,\cdots,k\,.  \eea For
$\mathbb{F}^{k-1}$, $\mathrm{dim}\ H_{2}(\mathbb{F}^{k-1},\R)=k+1$ and the
natural
basis is given by $\{l_{1},l_{2},e_{1},\dots, e_{k-1}\}$ (the $l_i$
come from $\PP^1 \times \PP^1$ and the $e_j$ are the blown-up $\PP^1$'s)
satisfying
\bea ^{}l_{i}\cdot l_{j}=1-\delta_{ij}\,,\,\,^{}l_{i}\cdot
e_{a}=0\,,\,\,^{}e_{a}\cdot
e_{b}=-\delta_{ab}\,,\,\,i,j=1,2\,;\,\,a,b=1,\cdots,k-1\,.  \eea Since
$\B_k \iso \FF^{k+1}$ we can write the two bases
just described in terms of one another; the map is given
by 
\begin{equation}
\begin{aligned}
 H&\mapsto l_{1}+l_{2}-e_{1}\,,\\
E_{1}&\mapsto l_{2}-e_{1}\,,\\ 
E_{2}&\mapsto l_{1}-e_{1}\,,\\
E_{a+1}&\mapsto e_{a}\,,\,\,\,\,a=2,\cdots,k-1\,. 
\end{aligned}
\end{equation}
The inverse map
is then
\begin{equation}
\begin{aligned} \label{homology-identification} 
l_{1} &\mapsto H-E_{1}\,,\\
l_{2}&\mapsto H-E_{2}\,,\\ 
e_{1}&\mapsto H-E_{1}-E_{2}\,,\\ 
e_{a}&\mapsto E_{a+1}\,,\,\,\,\,a=2,\cdots,k-1\,.
\end{aligned}
\end{equation}
The canonical class, defined to be
minus the first Chern class of the tangent bundle,
will play an important role in our correspondence to M-theory; it is given by
\begin{align}
K_{\mathbb{B}_{k}}=&-c_{1}(\mathbb{B}_{k})=-3H+\sum_{a=1}^{k}E_{a}\,,
\label{canonicalclass}\\
K_{\mathbb{F}^{k-1}}=&-c_{1}(\mathbb{F}^{k-1})=-2l_{1}-2l_{2}+\sum_{a=1}^{k-1}e_{a}\,,
\end{align}
 where $c_{1}(X)$ is the first Chern class of $X$.  Incidentally, from
(\ref{canonicalclass}) we see directly that the condition of positive first
Chern class is not satisfied for $\mathbb{B}_{k}$ or $\mathbb{F}^{k-1}$ with $k>8$, since
\bea
c_{1}^{2}(X)=K_X^2 = 9-k\,\,\,\,\mbox{for}\,\,\,X=\mathbb{B}_{k} \mathrm{\ or\ } X=\mathbb{F}^{k-1}\,.
\eea
We denote by $d_{{\cal C}}$ the degree of a class
${\cal C}\in H_{2}(\mathbb{B}_{k})$, which by definition is the intersection
of ${\cal C}$ with the anticanonical class, i.e. \bea \label{defdeg} d_{{\cal
C}}:=-{\cal C}\cdot K_{\mathbb{B}_{k}}\,.  \eea
The terminology ``degree'' is explained by the fact that one can use the anticanonical class
to give a map of $\B_k$ into projective space, and if $\cal C$ is represented by a holomorphic
curve then its image will have degree precisely $\degC$.

If the class ${\cal C}\in H_{2}(\mathbb{B}_{k})$ is realized as a
genus $g$ curve then we have an important relation between 
its self-intersection and its degree, known
as the adjunction formula \cite{MR95d:14001}:
\bea \label{adjunction}
^{}{\cal C}^2=2g-2+d_{{\cal C}}\,.
\label{ad}
\eea

\subsection{$E_{k}$ root lattice} We now 
review an important relation between del Pezzo
surfaces and exceptional Lie algebras.  Namely, the lattice
$H_{2}(\mathbb{B}_{k},\mathbb{Z})$ endowed with the intersection
product has a natural sublattice obtained by restricting
to the orthogonal complement of $K_{\mathbb{B}_{k}}$, which turns
out to be isomorphic to the root
lattice of $E_{k}$.  To see this (in case $k \ge 3$ --- for $k=1,2$ the
story is slightly irregular) 
a convenient basis for the sublattice
is the following, which is also a set of simple roots:
\begin{equation}
\begin{aligned}
\alpha_{i}&=E_{i}-E_{i+1}\,,\,\,\,\,i=1,\cdots,k-1\,,\\
\alpha_{k}&=H-E_{1}-E_{2}-E_{3}\,. 
\end{aligned}
\end{equation}
It is easy to see that these
classes are orthogonal to the canonical class, and that their intersection
numbers are given by
\bea K_{\mathbb{B}_{k}}\cdot \alpha_{a}=0\,,\,\,\, \label{cartan-matrix}
^{}\alpha_{a}\cdot
\alpha_{b}=-A_{ab}\,,\,\,a,b=1,\cdots,k\,,  \eea where
$A_{ab}$ represents the Cartan matrix of the Lie algebra $E_{k}$.  The Dynkin
diagrams of the Lie algebras $E_k$ are shown in \figref{ek}.
\onefigure{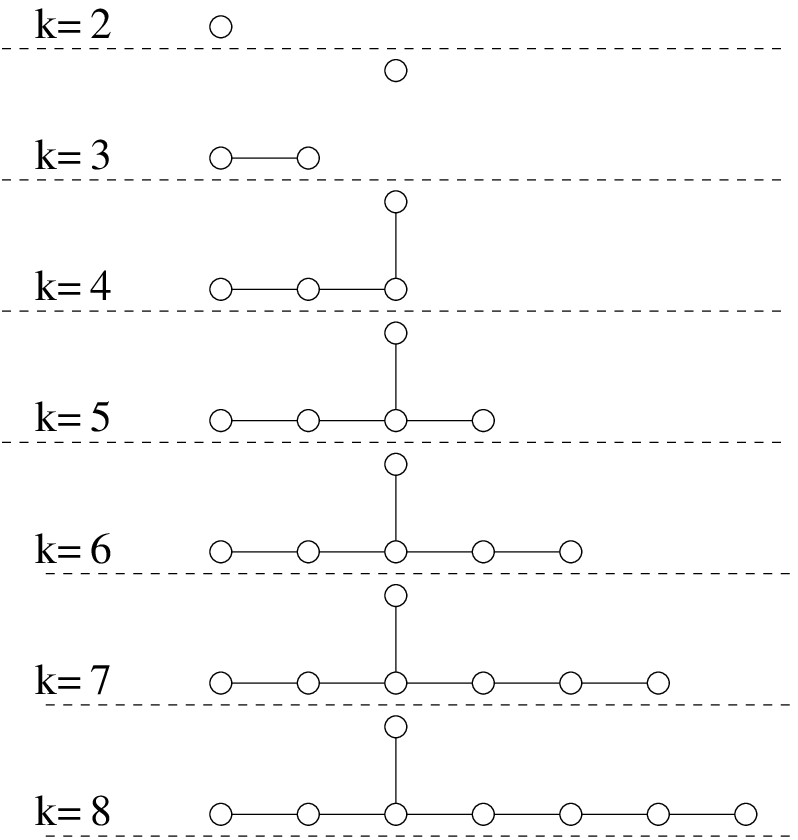}{The Dynkin diagrams of exceptional Lie algebras $E_{k}$.}
To each
${\cal C}\in H_{2}(\mathbb{B}_{k},\mathbb{Z})$ we can associate
an $E_{k}$ weight vector $\CC_\perp$,
obtained by projecting onto the orthogonal complement of $K_{\mathbb{B}_{k}}$.
The Dynkin labels of $\CC_\perp$ are given by
\bea \lambda_{a}=-^{}{\cal C}\cdot \alpha_{a}\,.  \eea In terms of
the weight vector, the self-intersection of ${\cal
C}\in H_{2}(\mathbb{B}_{k},\mathbb{Z})$ is given by \bea {\cal
C}^2=\CC_\perp^2+\frac{d_{{\cal C}}^{2}}{9-k}\,. \eea
Because of the minus sign in \eqref{cartan-matrix} we will have
$\CC_\perp^2 \le 0$.
Thus the lattice $H_{2}(\mathbb{B}_{k},\mathbb{Z})$ has signature $(1,k)$,
\bea
H_{2}(\mathbb{B}_{k},\mathbb{Z})=\IP{K_{\mathbb{B}_{k}}}\,\oplus \,\Gamma_{E_{k}}\,.
\eea
Here $\Gamma_{E_{k}}$ is the root lattice of $E_{k}$, with the sign
of the inner product reversed.

\subsection{Weyl group} \label{weyl} Next we show that the Weyl group of $E_{k}$ acts
naturally on $H_{2}(\mathbb{B}_{k}, \R)$, preserving the intersection
form and $K_{\mathbb{B}_{k}}$.  Given any $\alpha \in H_2(\B_k, \Z)$ with
$\alpha^2=-2$ and
$K_{\mathbb{B}_{k}}\cdot \alpha=0$ (a root) 
we can define a transformation $w_\alpha$, which
is the reflection in $\alpha$ and hence acts orthogonally: \bea
w_{\alpha}:\,\CC\mapsto \CC+^{}(\CC\cdot
\alpha)\,\alpha\,,\,\,\,\CC\in H_{2}(\mathbb{B}_{k})\,.  \eea

The elements $w_\alpha$ corresponding to simple roots have a particularly
interesting action on $H_{2}(\mathbb{B}_{k}, \R)$. The simple root
$\alpha_{i}=E_{i}-E_{i+1}$ exchanges $E_{i}$ and $E_{i+1}$
(we will see in a later section that this corresponds to
exchanging two of the circles on which we compactify M-theory.)
The action of
$\alpha_{k} = H - E_1 - E_2 - E_3$ 
is more nontrivial; on $\CC=nH-\sum_{a=1}^{k}m_{a}E_{a}$ it is
given by
\begin{equation}
\begin{split}
w_{\alpha_{k}}(\CC) =\ &(2n-m_{1}-m_{2}-m_{3})H-(n-m_{2}-m_{3})E_{1}-(n-m_{1}-m_{3})E_{2} \\
                      &-(n-m_{1}-m_{2})E_{3}-\sum_{a>3}m_{a}E_{a}\,.
\end{split}
\end{equation}
Note
that this transformation can only be defined for $k\geq 3$. We
will see later that it corresponds to a double T-duality, which from the 
11-dimensional point of view
requires three compact directions.  

The $w_{\alpha_i}$ together generate the Weyl group
of $E_k$; in fact this is the group of all automorphisms of $H_2(\B_k, \Z)$
which preserve the intersection form and $K$ \cite{MR87d:11037}.

This action of the Weyl group on $H_2(\B_k, \Z)$ can actually be 
realized by global diffeomorphisms
which act on the del Pezzo surface exchanging the exceptional curves and
preserving $K$.
For the roots $\alpha_i$ with $i < k$ this is easy to see:  we only have to take
a diffeomorphism of $\PP^2$ which exchanges two of the blown-up points while fixing
the rest.  Such diffeomorphisms certainly exist and can be extended to the full
del Pezzo.  What is more interesting is the reflection in $\alpha_k$:  this action is realized
by a mechanism which is less obvious from the perspective of $\PP^2$, 
which we discuss in Section \ref{cremona-section}.

Since the Weyl group action preserves $K$ it preserves the degree of any curve.
Thus curves of a given degree form a representation of the Weyl group.

\subsection{Rational curves} In our correspondence it will be especially important to 
understand the genus zero curves, also called rational curves, in
$H_{2}(\mathbb{B}_{k},\mathbb{Z})$.
Such a curve satisfies the adjunction formula
\eqref{adjunction} with $g=0$.  Hence if ${\cal
C}=nH-\sum_{a=1}^{k}m_{a}E_{a}$ is represented by a rational curve we have
\begin{equation}
\CC^2 = d_{{\cal C}}-2\,,
\end{equation}
or equivalently
\begin{equation}
n(n-3)-\sum_{a=1}^{k}m_{a}(m_{a}-1)=-2\,.
\label{rational}
\end{equation}
For example, classes in
$H_{2}(\mathbb{B}_{8},\mathbb{Z})$ which have $\degC = 1$ and satisfy \eqref{rational} are \cite{MR87d:11037}
\begin{equation}
\label{zerob}
\begin{aligned}
E_{a}\,,\,H-E_{a}-E_{b}\,,\,2H-\sum_{i=1}^{5}E_{a_{i}}\,,\,3H-2E_{a}-\sum_{i=1}^{6}E_{a_{i}}\,,\\
4H-2E_{a}-2E_{b}-2E_{c}-\sum_{i=1}^{5}E_{a_{i}}\,,5H-2\sum_{i=1}^{6}E_{a_{i}}-\sum_{j=1}^{2}E_{a_{j}}\,,\\
6H-3E_{a}-2\sum_{i=1}^{7}E_{a_{i}}\,.
\end{aligned}
\end{equation}
In terms of the
$E_{k}$ weight vector $\CC_\perp$ of a rational curve ${\cal
C}$ we have \bea \CC_\perp^2=-\frac{d^{2}_{{\cal
C}}}{9-k}+d_{{\cal C}}-2\,. \label{cperp} \eea
From \eqref{cperp} we see that the degree $1$ rational curves
(exceptional curves)
transform under the Weyl group in the fundamental representation
of $E_k$, since for $d_{\cal C}=1$ we would have
\bea
\CC_\perp^2=-\frac{10-k}{9-k}\,,
\eea
which is just what we would expect for weight vectors of the fundamental.

\subsection{Toric geometry} Toric geometry provides an interesting way
to visualize some of the del Pezzo surfaces as well as a neat diagrammatic method
of obtaining
the intersection numbers of curves on
toric del Pezzo surfaces.  We therefore give a review of the relevant aspects
of toric geometry applied to del Pezzo surfaces.
\medskip

\underline{$\PP^1$}:
We start with a simple example:  the representation of $\mathbb{P}^{1}$
in toric geometry, following the discussion in \cite{Leung:1998tw}. Since
$\mathbb{P}^{1}=S^{3}/U(1)$, we can
describe it as \cite{Witten:1993yc}
\begin{equation} \label{p1}
|\Phi_{1}|^{2}+|\Phi_{2}|^{2}=r\,,
\end{equation}
\begin{equation} \label{pu1}
U(1):\,\,(\Phi_{1},\Phi_{2}) \sim (\Phi_{1}e^{i\theta},\Phi_{2}e^{i\theta})\,.
\end{equation}
The complex variables $\Phi_{i}$ are related to the projective
coordinates $[z_{1},z_{2}]$ of $\PP^{1}$ by \bea
\Phi_{i}=r^{1/2}\,\frac{z_{i}}{\sqrt{\sum_{i}|z_{i}|^{2}}}\,. \eea The
geometry defined by (\ref{p1}) can be understood in a slightly
different way as well, which will be important for us when
discussing toric del Pezzo surfaces.  Namely, rewrite \eqref{p1}
and \eqref{pu1}
as follows: \bea \label{pp1} \label{redone-p1}
|\Phi_{2}|^{2}=r-|\Phi_{1}|^{2}\,,\,\,(\Phi_{1},\Phi_{2})\sim
(\Phi_{1}e^{i\theta},\Phi_{2}e^{i\theta})\,.  \eea Thus we see
that $0\leq |\Phi_{1}|^2\leq r$ and therefore the range of
$|\Phi_{1}|^2$ is an interval with endpoints given by
$|\Phi_{1}|^2=0,r$.  Fibered over every point of the interval except
the endpoints we have two circles, given by the phases of
$\Phi_{1}$ and $\Phi_2$, the magnitudes being fixed by
\eqref{pp1}. However, because of the $U(1)$ identification only
the relative phase survives; at every point of the interval except
the endpoints we have a finite size circle parametrized by this
phase. At one of the endpoints, say $|\Phi_{1}|=0$, we have
$|\Phi_{2}|^2=r$ with the phase arbitrary; however, this phase is
completely fixed by the $U(1)$ quotient.  Thus at the endpoints of
the interval the circle fiber shrinks to zero radius.  As shown in
\figref{toricp1} the total geometry is topologically a
$\mathbb{P}^{1}$. \onefigure{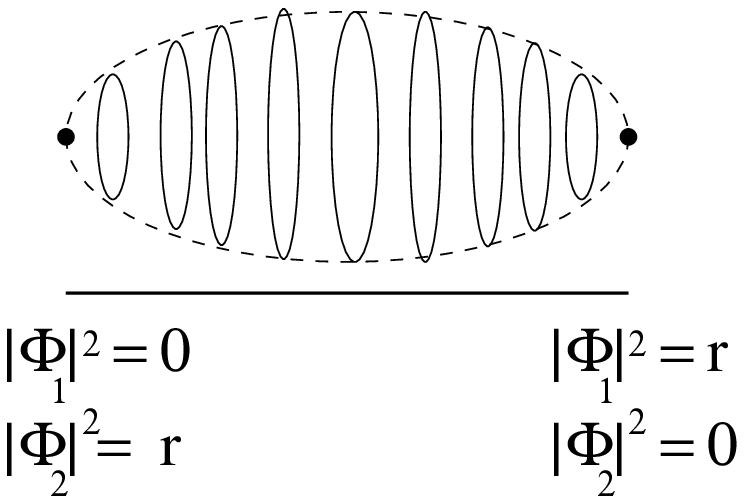}{$\mathbb{P}^{1}$ as a circle
fibration over an interval.}

\underline{$\mathbb{P}^{2}$}:\,
Now that we understand the representation of $\mathbb{P}^{1}$ as a
circle fibration over an interval, we can try to understand 
the analogous picture for toric
del Pezzo surfaces, starting with $\mathbb{P}^{2}$. In terms of complex
variables $\Phi_{1},\Phi_{2}$ and $\Phi_{3}$, $\mathbb{P}^{2}$ can be
represented as \cite{Witten:1993yc}
\begin{equation}
|\Phi_{1}|^{2}+|\Phi_{2}|^{2}+|\Phi_{3}|^{2}=r\,,
\label{p2toric}
\end{equation}
\begin{equation}
(\Phi_{1},\Phi_{2},\Phi_{3})\,\,\sim\,\,(\Phi_{1}e^{i\theta},\Phi_{2}e^{i\theta},\Phi_{3}e^{i\theta})\,.
\end{equation}
The complex variables are related to the projective coordinates
$[z_{1},z_{2},z_{3}]$ of $\PP^{2}$, \bea
\Phi_{i}=\sqrt{r}\frac{z_{i}}{\sqrt{\sum_{i=1}^{3}|z_{i}|^{2}}}\,.
\eea The equtaion (\ref{p2toric}) defines an $S^{5}$ of radius $\sqrt{r}$, as
can be seen easily by writing the equation in terms of the real
and imaginary parts of the $\Phi_{i}$. The $U(1)$ identification
then gives us $\mathbb{P}^{2}$ as $S^{5}/U(1)$ with K\"ahler
parameter $r$. In analogy with the $\mathbb{P}^{1}$ case we then
see that the toric representation of $\mathbb{P}^{2}$ will be an
$S^{1}\times S^{1}$ fibration over some base, with the
$S^{1}\times S^{1}$ given by the relative phases among
$\Phi_{1},\Phi_{2},\Phi_{3}$.

To obtain the base of the fibration we rewrite the equation defining
$S^{5}$ as \bea \label{redone-p2}
|\Phi_{3}|^{2}=r-|\Phi_{1}|^{2}-|\Phi_{2}|^{2}\,.  \eea Since
$|\Phi_{3}|\geq 0$ the base
is a triangular region in the plane, with
coordinate axes parametrized by $|\Phi_{1}|^{2}$ and $|\Phi_{2}|^{2}$ and
boundary given by three intervals as shown in
\figref{toricp2}(a),
\begin{equation}
\begin{aligned} \label{pp2}
I_{1}:&\,|\Phi_{1}|^{2}=0,\,\,0\leq |\Phi_{2}|^{2}\leq r\,,\\
I_{2}:&\,|\Phi_{2}|^{2}=0\,,\,\,0\leq |\Phi_{1}|^{2}\leq r\,,\\
I_{3}:&\,|\Phi_{1}|^{2}+|\Phi_{2}|^{2}=r\,,\,\,|\Phi_{1}|,|\Phi_{2}|\geq
0\,.
\end{aligned}
\end{equation}
At every point {\it inside} the triangle the magnitudes of
$\Phi_{1},\Phi_{2}$ and $\Phi_{3}$ are fixed but the phases are not,
giving a $T^2$ for the relative phases.  At the boundary the situation is different
since some of the $\Phi_{i}$ are zero and hence the corresponding
circles have collapsed. From
(\ref{pp2}) we see that at the boundary
component $I_{i}$ we have $\Phi_{i}=0$. When one of the $\Phi_{j}$ vanishes
\eqref{redone-p2} 
reduces to \eqref{redone-p1}; thus the interval represents
a $\mathbb{P}^{1}$ inside $\mathbb{P}^{2}$, as shown in
\figref{toricp2}(b). At points where two of the $I_i$ intersect
(the vertices of the triangle) only one $\Phi_{i}$ is non-zero;
its magnitude is determined by \eqref{redone-p2}, and the
phase is completely fixed by the $U(1)$ quotient.
Thus we see that inside the triangle we have a $T^{2}$
fibration, at the edges the $T^{2}$ collapses to an $S^{1}$,
and at the vertices the $S^{1}$ collapses to a point. The
three intervals $I_{i}$ represent three $\mathbb{P}^{1}$'s.
\onefigure{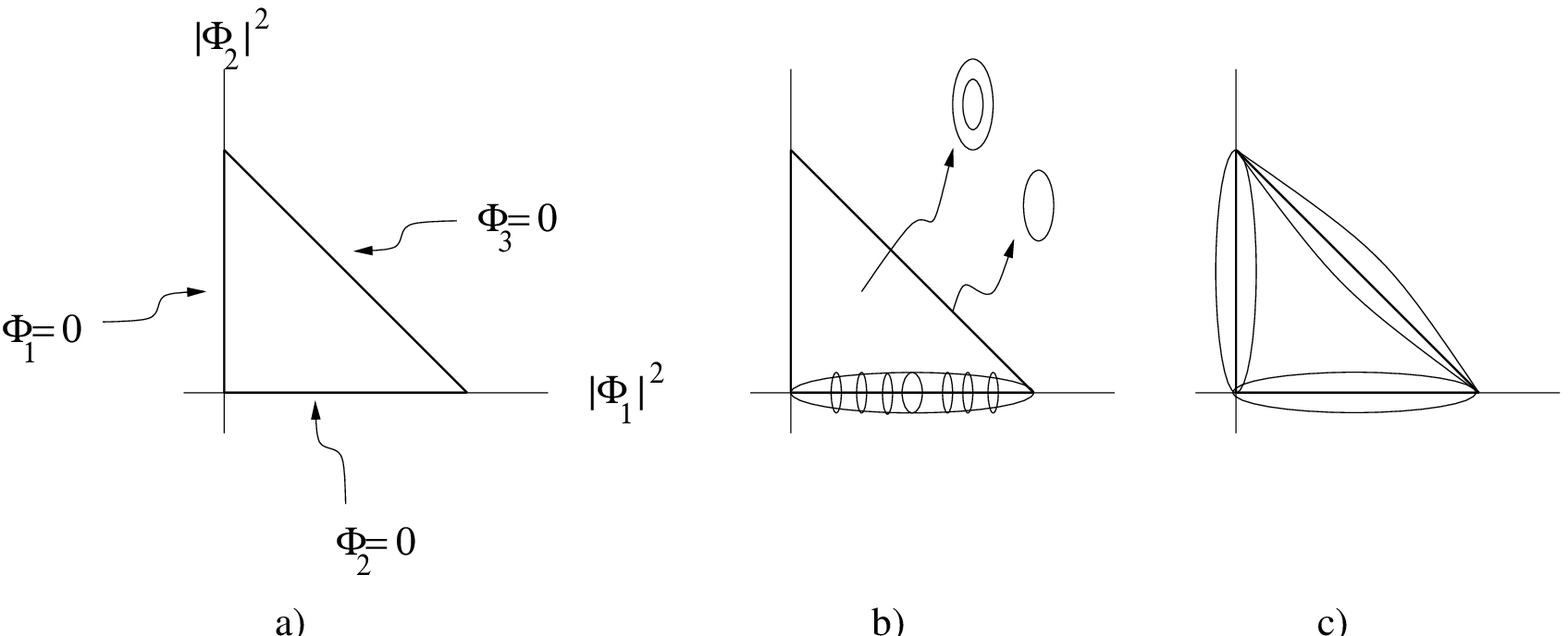}{a) The toric diagram of $\mathbb{P}^{2}$, b) $\mathbb{P}^{2}$ as a $T^{2}$ fibration with collapsing fibers at the boundary, c) The three boundary components as $\mathbb{P}^{1}$'s.}

\medskip
\underline{$\mathbb{B}_{1}$}:\,
Next we consider the case of $\mathbb{P}^{2}$ blown up at one point.
In this case we have four complex variables
$\Phi_{1},\Phi_{2},\Phi_{3}$ and $\Phi_{4}$, satisfying
\begin{equation} \label{b1a}
\begin{aligned}
|\Phi_{1}|^{2}+|\Phi_{2}|^{2}+|\Phi_{3}|^{2}&=r_{1}\,,\\
|\Phi_{1}|^{2}+|\Phi_{4}|^{2}&=r_{2}\,,\,\,r_{1}\geq r_{2}\geq 0\,,\\
\end{aligned}
\end{equation}
\begin{equation}
\begin{aligned}
U(1)_{a}&:\,(\Phi_{1},\Phi_{2},\Phi_{3},\Phi_{4})\sim
(e^{i\theta_{a}}\Phi_{1},e^{i\theta_{a}}\Phi_{2},e^{i\theta_{a}}\Phi_{3},\Phi_{4})\,,\\
U(1)_{b}&:\,(\Phi_{1},\Phi_{2},\Phi_{3},\Phi_{4})\sim
(e^{i\theta_{b}}\Phi_{1},\Phi_{2},\Phi_{3},e^{i\theta_{b}}\Phi_{4})
\label{b1eq}
\end{aligned}
\end{equation}
The first equation in \eqref{b1a} defines a $\mathbb{P}^{2}$
and the second defines a $\mathbb{P}^{1}$; $r_{1}$ and $r_{2}$ are the two K\"ahler parameters of the surface. We can draw the
toric picture in the plane, parametrized as before by
$|\Phi_{1}|^{2}$ and $|\Phi_{2}|^{2}$. Eliminating $|\Phi_{3}|^{2}$
and $|\Phi_{4}|^{2}$ we get linear constraints on $|\Phi_{1}|^{2}$ and
$|\Phi_{2}|^{2}$ which define the base of the torus fibration: 
\begin{equation}
\begin{aligned}
|\Phi_{1}|^{2}+|\Phi_{2}|^{2}&\leq r_{1}\,,\\
|\Phi_{1}|^{2}&\leq r_{2}\,.
\end{aligned}
\end{equation}
So now we have four
boundary components, given by
\begin{equation}
\begin{aligned}
I_{1}:&\,|\Phi_{1}|^{2}=0\,,\,\,0\leq |\Phi_{2}|^{2}\leq r_{1}\,,\\
I_{2}:&\,|\Phi_{2}|^{2}=0\,,\,\,0\leq |\Phi_{1}|^{2}\leq
r_{2}\,,\\
I_{3}:&\,|\Phi_{1}|^{2}=r_{2}\,,\,0\leq
|\Phi_{2}|^{2}\leq r_{1}-r_{2}\,,\\
I_{4}:&\,0\leq
|\Phi_{1}|^{2}\leq r_{2}\,,\,r_{1}-r_{2}\leq |\Phi_{2}|^{2}\leq
r_{1}\,.
\end{aligned}
\end{equation}
The region bounded by the $I_i$ is
the base of the fibration, shown in \figref{toricb1-b}.
\onefigure{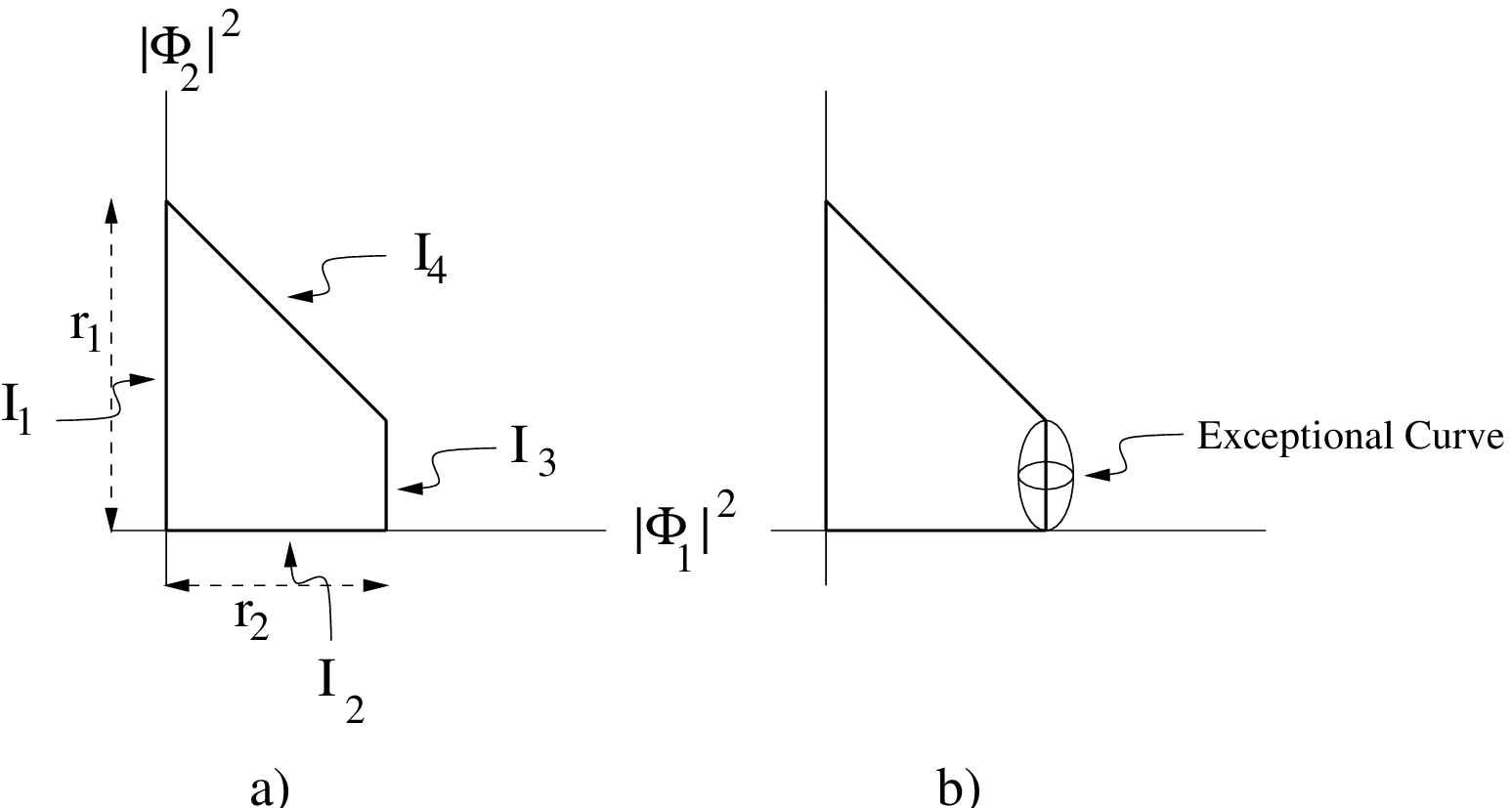}{a) The toric diagram of $\mathbb{B}_{1}$. b) The
interval which replaces the rightmost vertex of the triangle
represents the exceptional curve.}
If we had used $|\Phi_{2}|^{2}$ in the second line of (\ref{b1a})
then a different vertex of the triangle would have been replaced by a line segment; the choice among $|\Phi_{1}|^{2}$,
$|\Phi_{2}|^{2}$ and $|\Phi_{3}|^{2}$ corresponds in the
toric picture to the choice of which vertex to replace by
a line segment representing the exceptional curve.

\medskip
\underline{$\mathbb{B}_{2}$}:\,
Now let us consider the case of $\mathbb{P}^{2}$ blown up at two
points. The equations representing $\B_2$ are
\begin{equation}
\begin{aligned} \label{a1}
|\Phi_{1}|^{2}+|\Phi_{2}|^{2}+|\Phi_{3}|^{2}&=r_{1}\,,\\
|\Phi_{1}|^{2}+|\Phi_{4}|^{2}&=r_{2}\,,\\
|\Phi_{2}|^{2}+|\Phi_{5}|^{2}&=r_{3}\,,\,\,\,\,\,r_{1}\geq r_{2}, r_1 \ge r_3\,,
\end{aligned}
\end{equation}
\begin{equation}
\begin{aligned} \label{a2}
U(1)_{a}&:
(\Phi_{1},\Phi_{2},\Phi_{3},\Phi_{4},\Phi_{5})\sim
(e^{i\theta_{a}}\Phi_{1},e^{i\theta_{a}}\Phi_{2},e^{i\theta_{a}}\Phi_{3},\Phi_{4},\Phi_{5})\,,\\
U(1)_{b}&:(\Phi_{1},\Phi_{2},\Phi_{3},\Phi_{4},\Phi_{5})\sim
(e^{i\theta_{b}}\Phi_{1},\Phi_{2},\Phi_{3},e^{i\theta_{b}}\Phi_{4},\Phi_{5})\,,\\
U(1)_{c}&:(\Phi_{1},\Phi_{2},\Phi_{3},\Phi_{4},\Phi_{5})\sim
(\Phi_{1},e^{i\theta_{c}}\Phi_{2},\Phi_{3},\Phi_{4},e^{i\theta_{c}}\Phi_{5})\,.
\end{aligned}
\end{equation}
In the above equations $r_{1},r_{2}$ and $r_{3}$ are the three K\"ahler parameters
of the surface $\mathbb{B}_{2}$. As before, we solve these equations in the plane parametrized by $|\Phi_{1}|^{2}$ and
$|\Phi_{2}|^{2}$. Eliminating $|\Phi_{3}|^{2},|\Phi_{4}|^{2}$ and
$|\Phi_{5}|^{2}$ leads to inequalities in terms of $|\Phi_{1}|^{2}$
and $|\Phi_{2}|^{2}$ whose solutions bound a region in the
plane, with boundary components given by 
\begin{equation}
\begin{aligned}
I_{1}:&
|\Phi_{1}|^{2}=0\,,\,0\leq |\Phi_{2}|^{2}\leq r_{3}\,,\\
I_{2}:& |\Phi_{2}|^{2}=0\,,\,0\leq |\Phi_{1}|^{2}\leq r_{2}\,,\\
I_{3}:& |\Phi_{1}|^{2}=r_{2}\,,\,0\leq |\Phi_{2}|^{2}\leq
r_{1}-r_{2}\,,\\ I_{4}:& r_{1}-r_{3}\leq |\Phi_{1}|^{2}\leq
r_{2}\,,\,r_{1}-r_{2}\leq |\Phi_{2}|^{2}\leq
r_{3}\,,|\Phi_{1}|^{2}+|\Phi_{2}|^{2}=r_{1}\,,\\ I_{5}:&
|\Phi_{2}|^{2}=r_{3}\,,\,0\leq |\Phi_{1}|^{2}\leq r_{1}-r_{3}\,.
\end{aligned}
\end{equation}
The base of the torus fibration is
shown in \figref{toricb2}.  \onefigure{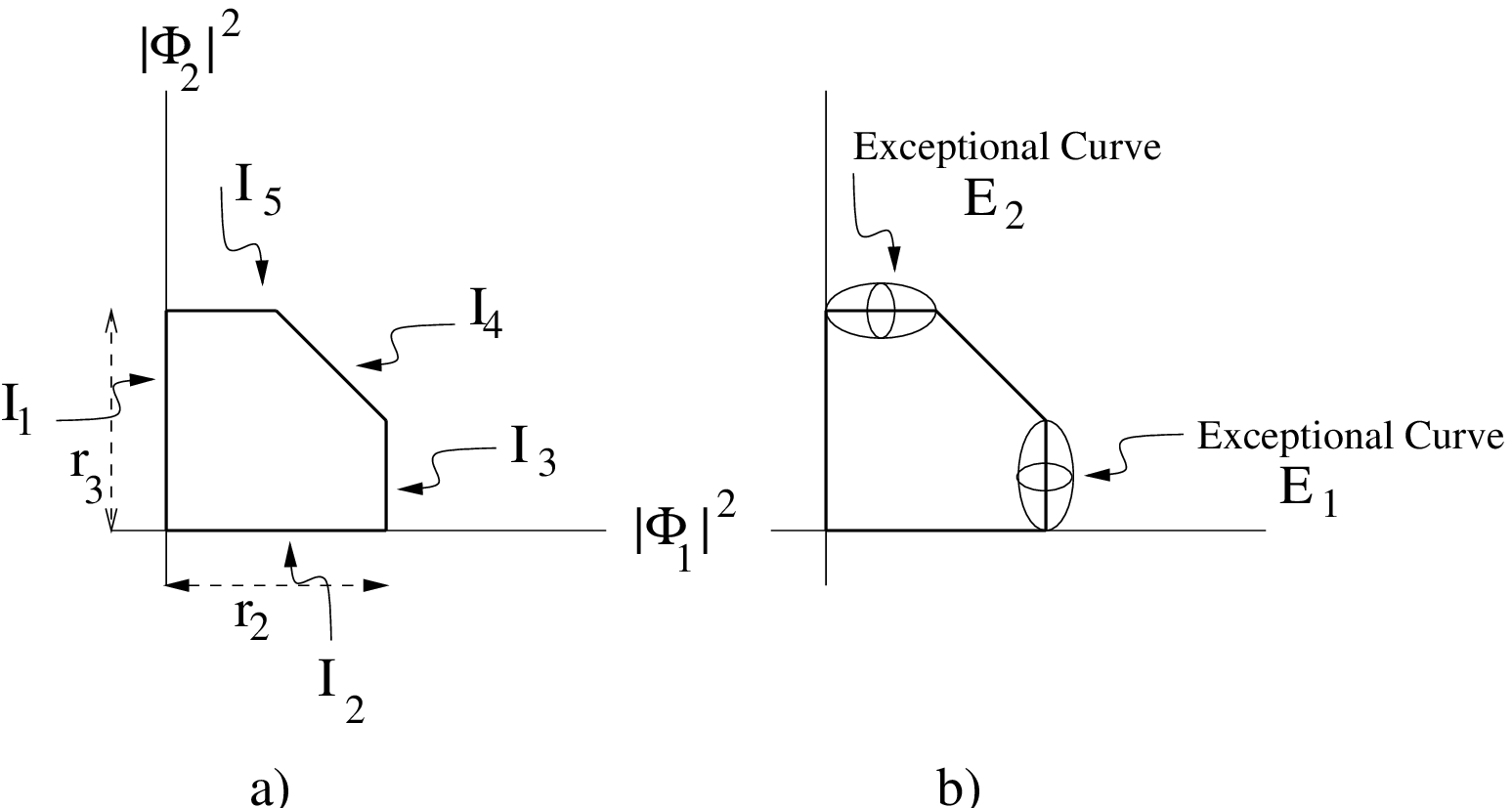}{a) The toric diagram of $\mathbb{B}_{2}$, b) The two intervals replacing the vertex represent the two exceptional curves on the base.}

\medskip
\underline{$\B_3$}:  Next we consider $\PP^2$ blown up at three points.
We omit the equations but they are similar to those in the previous examples; there are three exceptional curves now, described
torically by three intervals which replace the vertices of the triangle.  The picture is
shown in \figref{toricb3}.  \onefigure{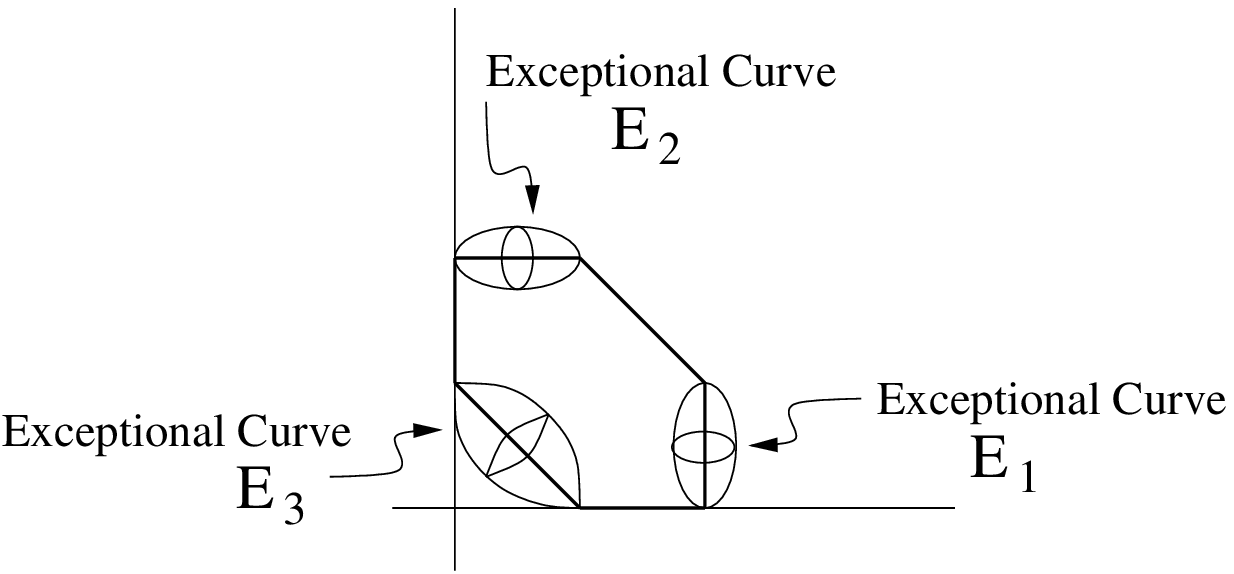}{The toric diagram of $\mathbb{B}_{3}$ with three exceptional curves as intervals on the base.}

\medskip
From the diagrams we have seen so far it is clear that the toric
representation of the operation of blowing up a point which is a
vertex of the toric diagram is the replacement of that point by a
line segment.  In case $k \le 3$ any triplet of points on $\PP^2$ is equivalent
in the sense that there is an automorphism of $\PP^2$ which moves any given triplet to
the three vertices of the triangle; hence it is always sufficient to consider blowing
up the vertices.  For $k>3$ we encounter a difficulty:  after blowing up three points
all the curves appearing on the boundary of the toric diagram are exceptional curves, hence
rigid.  A generic point does not lie on any of these exceptional curves, so we cannot give a
toric description of the process of blowing up a fourth point; hence $\B_k$ does not
admit a toric description for $k>3$.

{}From the toric
diagrams we can easily see why $\mathbb{P}^{2}$ blown up at two points
is the same as $\mathbb{P}^{1}\times \mathbb{P}^{1}$ blown up at one
point, in other words $\B_2 \iso \FF^1$.
To see this consider first the case of $\mathbb{P}^{1}\times
\mathbb{P}^{1}$, which is given by the equations:
\begin{equation}
\begin{aligned}
|\Phi_{1}|^{2}&+|\Phi_{3}|^{2}=s_{1}\,,\\
|\Phi_{2}|^{2}&+|\Phi_{4}|^{2}=s_{2}\,,
\end{aligned}
\end{equation}
\begin{equation}
\begin{aligned}
U(1)_{a}:&(\Phi_{1},\Phi_{2},\Phi_{3},\Phi_{4})\sim
(e^{i\theta_{a}}\Phi_{1},\Phi_{2},e^{i\theta_{a}}\Phi_{3},\Phi_{4})\,,\\
U(1)_{b}:&(\Phi_{1},\Phi_{2},\Phi_{3},\Phi_{4})\sim
(\Phi_{1},e^{i\theta_{b}}\Phi_{2},\Phi_{3},e^{i\theta_{b}}\Phi_{4})\,.
\end{aligned}
\end{equation}
The toric diagram is shown in \figref{p1p1}.
\onefigure{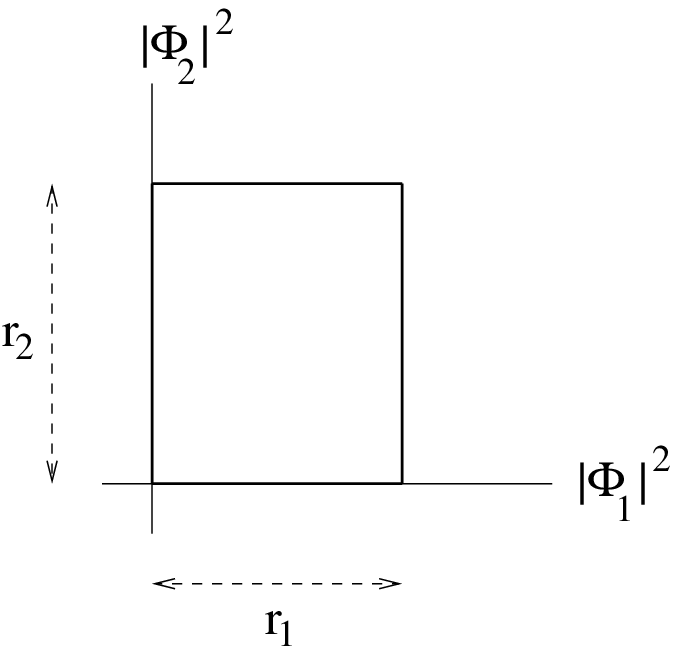}{The toric diagram of $\mathbb{P}^{1}\times \mathbb{P}^{1}$.}  Then we can consider $\mathbb{P}^{1}\times
\mathbb{P}^{1}$ blown up at one point, which is given by
\begin{equation}
\begin{aligned} \label{b1}
|\Phi_{1}|^{2}+|\Phi_{3}|^{2}&=s_{1}\,,\\
|\Phi_{2}|^{2}+|\Phi_{4}|^{2}&=s_{2}\,,\\
|\Phi_{1}|^{2}+|\Phi_{2}|^{2}+|\Phi_{5}|^{2}&=s_{3}\,,
\end{aligned}
\end{equation}
\begin{equation}
\begin{aligned} \label{b2}
U(1)_{a}:&(\Phi_{1},\Phi_{2},\Phi_{3},\Phi_{4},\Phi_{5})\sim
(e^{i\theta_{a}}\Phi_{1},\Phi_{2},e^{i\theta_{a}}\Phi_{3},\Phi_{4},\Phi_{5})\,,\\
U(1)_{b}:&(\Phi_{1},\Phi_{2},\Phi_{3},\Phi_{4},\Phi_{5})\sim
(\Phi_{1},e^{i\theta_{b}}\Phi_{2},\Phi_{3},e^{i\theta_{b}}\Phi_{4},\Phi_{5})\,,\\
U(1)_{c}:&(\Phi_{1},\Phi_{2},\Phi_{3},\Phi_{4},\Phi_{5})\sim
(e^{i\theta_{c}}\Phi_{1},e^{i\theta_{c}}\Phi_{2},\Phi_{3},\Phi_{4},e^{i\theta_{c}}\Phi_{5})\,.
\end{aligned}
\end{equation}
After a change of coordinates the equations \eqref{b1}, \eqref{b2} are equivalent to \eqref{a1},
\eqref{a2}, showing that $\B_2 \iso \FF^1$; but it is much easier to see this directly from
the toric diagram, as shown in
\figref{equiv}.
\onefigure{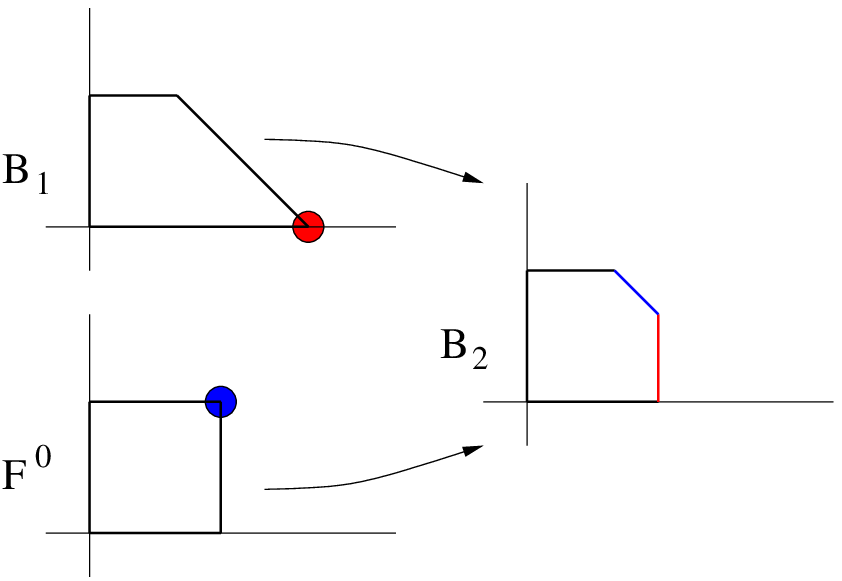}{Equivalence of $\mathbb{B}_{2}$ and $\mathbb{F}^{1}$.}

\subsection{Toric description of curves and intersection numbers}

As remarked earlier,
each boundary line segment of the toric diagram corresponds to an
exceptional curve. Let us denote the curve corresponding to
$I_{i}$ by $D_{i}$; then we can see from
\figref{curvesb3} that \bea
\sum_{i}D_{i}=-K\,,\,\,\,\mbox{and}\,\,\,^{}D_{i}\cdot
D_{j}=\delta_{i,j+1}+\delta_{i,j-1}\,.  \eea
\onefigure{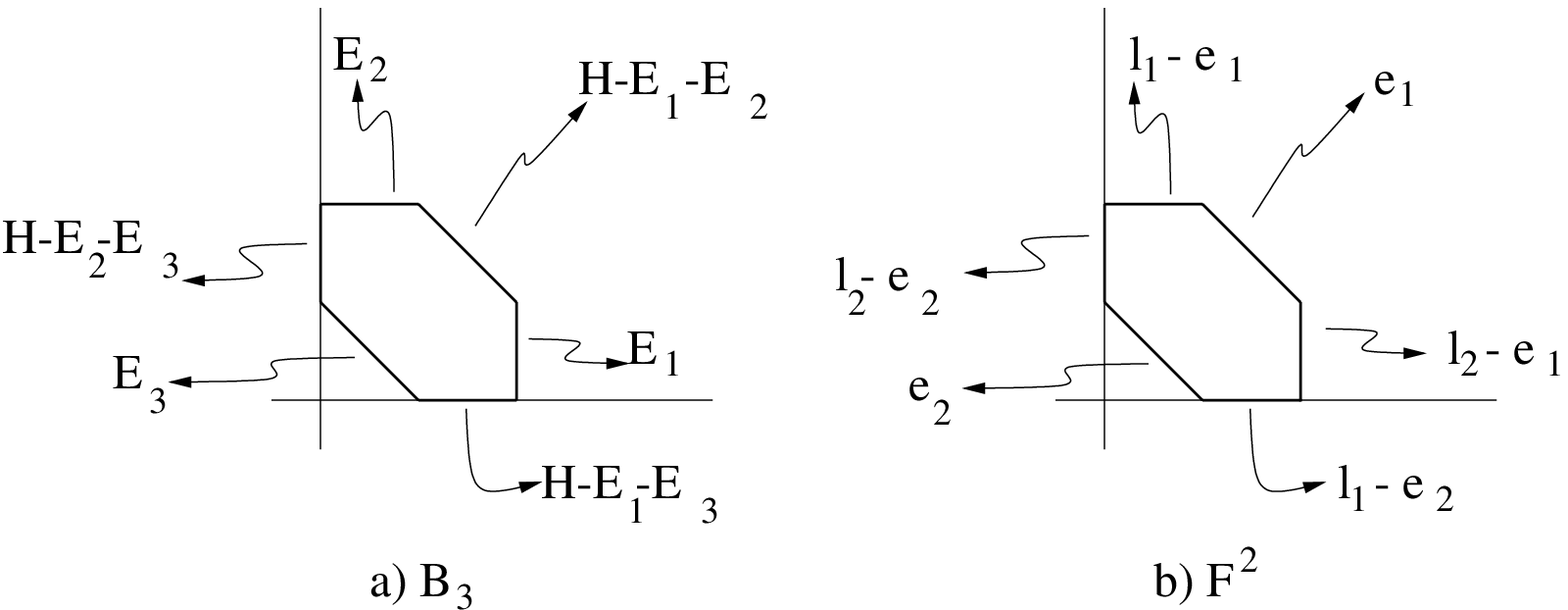}{a)
Boundary components of the toric diagram and corresponding exceptional
curves for $\mathbb{B}_{3}$, b) boundary components and corresponding
exceptional curves for $\mathbb{F}^{2}$.}

An equivalent way of representing curves in the del Pezzo is 
motivated from the string web picture in the context of $(p,q)$
5-branes
\cite{Aharony:1997ju},
which is related to del Pezzos as discussed in \cite{Leung:1998tw}.
Namely, the del Pezzos have a symplectic form represented
by the K\"ahler class. The base of the toric fibration can
be viewed as the `$x$' space and the $T^2$ fibers can be viewed
as `$p$' directions, where we represent the symplectic
form as $\sum_{i=1}^2dx_i\wedge dp_i=\sum_{i=1}^2 d|\Phi_i|^2 \wedge d\theta_i$,
where $\theta_i$ represents the phase of $\Phi_i$.  This implies
that each direction in the base is naturally paired with a circle in the fiber.
Consider for example a line in the base ending on a boundary. 
If the line ends orthogonally on the boundary, considering
the total space of the line and the corresponding circle in the fiber
 one obtains a piece of a holomorphic curve.  
We can connect these pieces together to obtain closed curves by
 drawing trivalent
graphs with external legs ending orthogonally on the boundary
components $I_{i}$. Consider first the case of
$\mathbb{P}^{2}$; let us try to represent in this way the class $H$ 
of a line.  If we can find a curve of genus zero with self-intersection
one then it is clear that this curve represents $H$.  As shown in
\figref{p2}(a), such a curve corresponds to the simplest trivalent
graph in the triangle; this can be understood from the fact that the
boundary components are in the same class as $H$ and therefore \bea
^{}D_{i}\cdot H=1\,, \,\,\,i=1,2,3,\eea which implies that the graph
corresponding to $H$ should have one external leg ending on each
$I_{i}$.   In fact note that
there is a moduli space for this curve
(which is again a $\PP^2$) which can be viewed in the
base as the choice for the position of the trivalent vertex
(this description
itself gives the base of a toric realization of
the moduli space).
One can see that if we go to a point on the moduli space
of this curve corresponding to putting the trivalent
point of the graph at one of the three vertices
of the triangle we obtain
the curve $z_1=0$ which represents $H$.

 \figref{p2}(b) shows the case of a conic in $\PP^2$, which
has genus zero and in homology is just $2H$. It is easy to see by
deforming the graph and making it trivalent that it indeed represents
a curve of genus zero.  On the other hand, a cubic curve will have
genus one, as one sees from the ``hole'' appearing in the rightmost
picture in \figref{p2}(c).  \onefigure{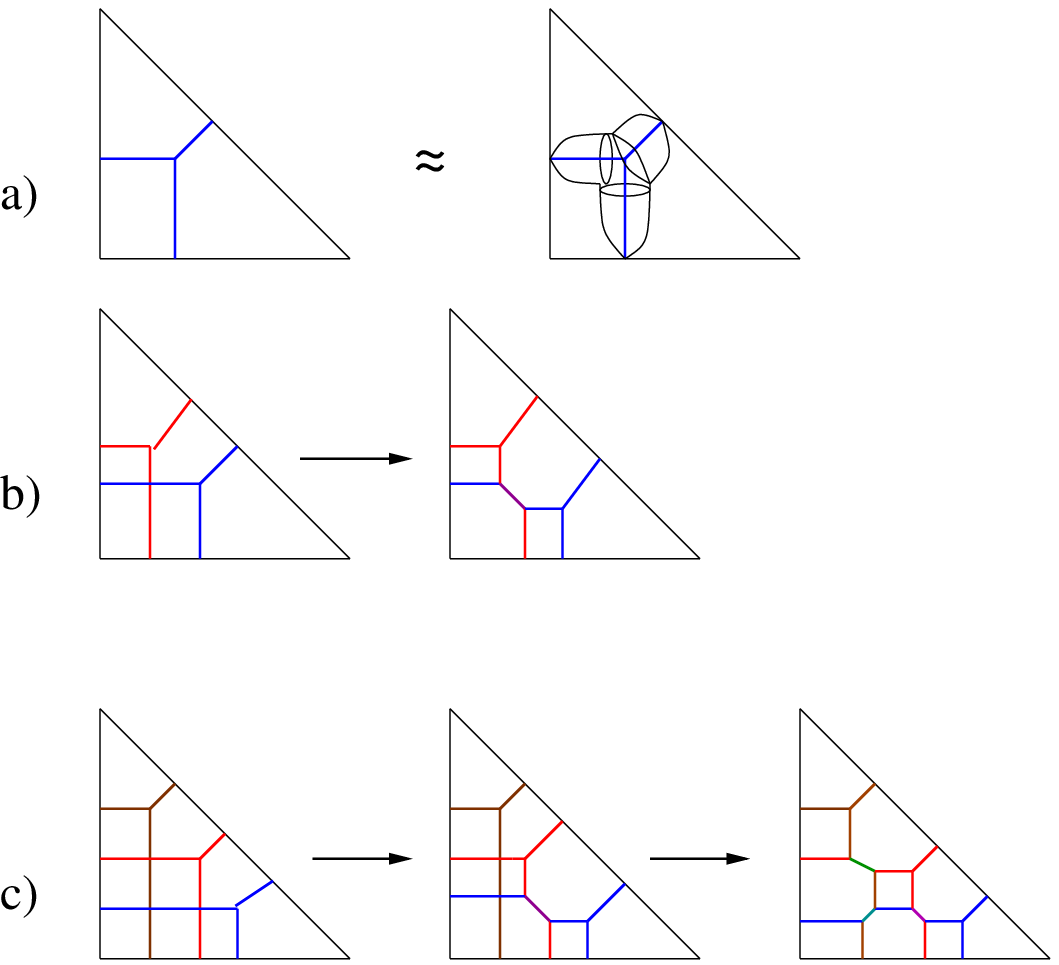}{a) Line $H$ in
$\mathbb{P}^{2}$, b) conic $2H$ in $\mathbb{P}^{2}$, c) anticanonical
class $3H$ in $\mathbb{P}^{2}$.}

\subsection{Quadratic transformation} \label{cremona-section}

Now we discuss the geometric realization of the nontrivial
part of the Weyl group, the reflection in $\alpha_k = H - E_1 - E_2 - E_3$.
First consider the case $k=3$ and 
observe that starting from $\mathbb{B}_{3}$ we can obtain
$\PP^2$ in two distinct ways: 
either blow down the set $\{E_{1},E_{2},E_{3}\}$ of exceptional curves, 
or blow down
$\{H-E_{2}-E_{3},H-E_{1}-E_{3},H-E_{1}-E_{2}\}$ to obtain
$\PP^{2}$, as shown in \figref{cremona}.
\onefigure{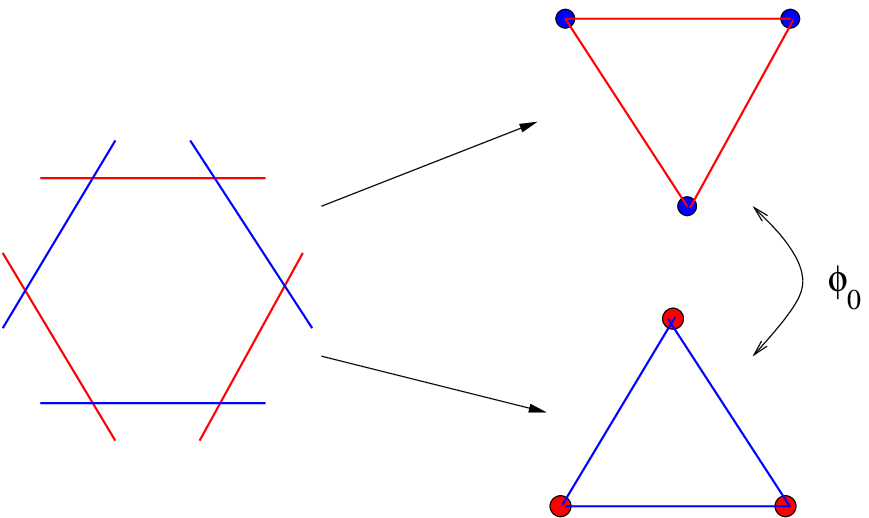}{Blowing down from $\mathbb{B}_{3}$ to
$\mathbb{P}^{2}$ in two different ways.}  The two sets of
exceptional curves are related to each other precisely by 
the reflection in the root $\alpha_{3}.$
Correspondingly, the $\PP^{2}$'s obtained this way are
related by a birational map $\phi_0$ called the ``quadratic transformation,'' 
\cite{MR95d:14001}
given by \bea \label{defquad}
\phi_{0}\,:\,[z_{1},z_{2},z_{3}]\,\mapsto\,[z_{2}z_{3},z_{1}z_{3},z_{1}z_{2}]\,.
\eea 
The expression above makes sense only when at least two of the $z_i$
are nonzero (otherwise we get $[0,0,0]$ which is not a point of
$\PP^2$); so $\phi_0$ is defined on $\PP^2$ minus the three points
$x_1 = [1,0,0], x_2 = [0,1,0], x_3 = [0,0,1]$.
By blowing up those three points to $E_1, E_2, E_3$ 
we can make $\phi_0$ defined everywhere.
Similarly, we can make $\phi_0$ 1-1 by blowing up the same three
points in the image $\PP^2$.  We therefore obtain an automorphism of $\B_3$.

Now how does $\phi_0$ act on $H_2(\B_3, \Z)$?
It is easy to see that \bea
\phi_{0}^{2}([z_{1},z_{2},z_{3}])=[(z_{1}z_{2}z_{3})\,z_{1},(z_{1}z_{2}z_{3})\,z_{2},(z_{1}z_{2}z_{3})\,z_{3}]\,.
\eea Thus $\phi_{0}^{2}$ is the identity on the open set
$\PP^{2}-\{z_{1}=0\}-\{z_{2}=0\}-\{z_{3}=0\}$ and hence extends to the
identity on $\B_3$, so $\phi_0^2 = 1$ on $H_2(\B_3, \Z)$.
Next let us consider the action of $\phi_0$ on $H - E_1 - E_2$.  This class
is the proper transform of a complex line in $\PP^2$ passing through the
points $x_1$ and $x_2$, namely the line $z_3 = 0$.  The action \eqref{defquad}
of $\phi_0$ maps this line to the point $x_3 = [0,0,1]$ which was blown up
to obtain $E_3$; hence $\phi_0(H - E_1 - E_2) = E_3$.  Since $\phi_0^2 = 1$
we also have $\phi_0(E_3) = H - E_1 - E_2$ and similarly for the other 
exceptional curves.  This is precisely the action of the reflection in
the root $\alpha_3$, as desired.

For $k > 3$ the situation is similar:  the presence of additional blown-up
points at generic positions on $\PP^2$ does not change anything, except
that the quadratic transformation will move these points, so that we
have to compose with a diffeomorphism of $\PP^2$ 
to move them back to their original positions.  So for all
$k \ge 3$ the
reflection in $\alpha_k$ is realized by a global diffeomorphism.

The description of the quadratic transformation above was given in terms
of the realization of $\B_k$ as $\PP^2$ blown up at $k$ points.  However,
as discussed above, $\B_k \iso \FF^{k+1}$ so we can equally well think
of $\PP^1 \times \PP^1$ blown up at $k-1$ points.  From this perspective
the action of the quadratic transformation is simple to describe:  namely,
it corresponds to exchanging two of the $k-1$ blown-up points.  This 
description makes it obvious that the quadratic transformation is realized
by a global diffeomorphism.

\section{The correspondence}

 We have discussed in the previous section how the moduli
 spaces of del Pezzos are interrelated via the operations
of blowing up and down.  Moreover, as we have
discussed, the group of global diffeomorphisms of $\B_k$ which fix
$K$ is the Weyl group of the corresponding exceptional group $E_k$.

On the other hand, if we consider compactifications
of M-theory on rectangular tori $T^k$ with no vev for the $C$-field,
the U-duality group acting
on this class of compactifications
is also realized as the Weyl group of $E_k$ \cite{Elitzur:1998zn}
(see also \cite{Banks:1998vs}.)  It is thus natural to ask if there
is a map between such M-theory compactifications
and del Pezzo geometries.   In fact, the story is already rather interesting
for small $k$'s, as shown in
\figref{toric2}.

\onefigure{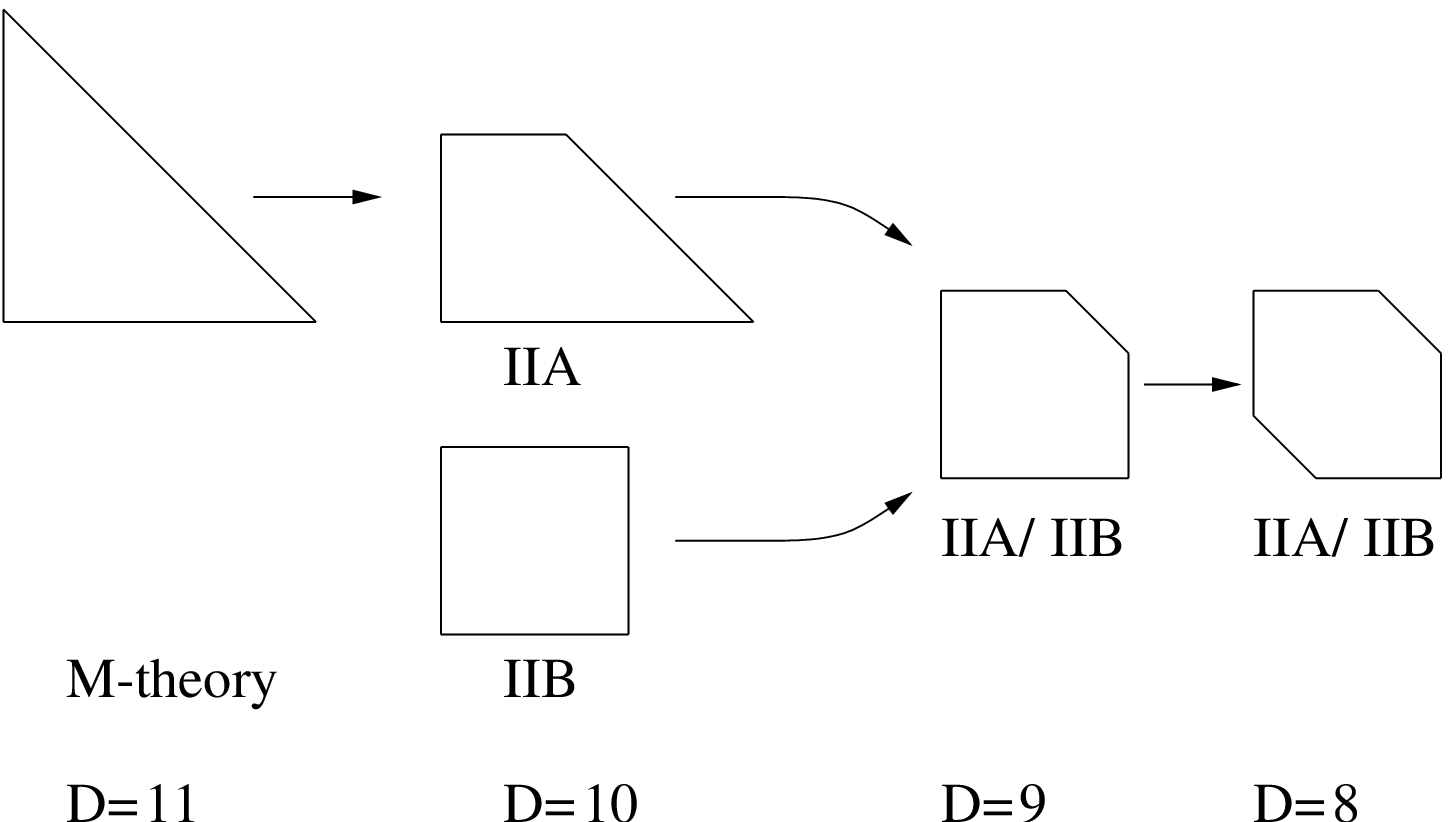}{M-theory / del Pezzo relation for $k=0,1,2,3$.}

Note that in \figref{toric2} each time we blow up a point to a $\PP^1$
in the del Pezzo, we compactify a circle on the M-theory side.  It is remarkable
that the exotic role type IIB plays in the chain of dualities of M-theory
exactly matches the role $\PP^1\times \PP^1$ plays among
del Pezzos.  In particular, if we wish to get from M-theory in 11-dimensions
to type IIB in 10 dimensions, we have to compactify two circles
and then let the resulting torus shrink to zero size; this is precisely mirrored by the fact
that to get to $\PP^1\times \PP^1$ from $\PP^2$ we first have
to blow up two points and then blow down another $\PP^1$.  Note
also that the symmetries of the corresponding theories are manifest
as classical symmetries of the del Pezzos.  For example, the $S$-duality
of type IIB corresponds to the exchange of the two sides of the rectangle
above (i.e. the exchange of the two $\PP^1$'s).  Moreover, the
non-trivial part of S-duality
in M-theory compactified to 9 dimensions is still $\Z_2$, which is reflected
in the $\Z_2$ symmetry of the rectangle with one corner cut off.  The non-trivial
part of S-duality in $d=8$ is $\Z_2\times S_3$,
which is manifest as the symmetries of the del Pezzo $\B_3$, torically
represented by the group of symmetries of the hexagon depicted
in the above figure.  Indeed this identification of Weyl groups
continues to make sense for all
del Pezzos, even those which are not toric (namely the
$\B_k$ with $k>3$).

It is thus natural to expect that we can get a more detailed map
between the two sides and promote the above correspondence to the level
of moduli and objects on both sides.  We now consider this question.

\subsection{Matching the moduli spaces} \label{moduli-spaces}

We consider the moduli space for M-theory
on a geometry $T^k \times \R^{10-k,1}$ for
$k \le 8$, where the torus has a rectangular metric of the form
\begin{equation}
ds^2 = \sum_{i=1}^k R_i^2 dx_i^2
\end{equation}
with the $x_i$ periodic, $x_i \sim x_i + 2 \pi$,
and all background $C$ fields set to zero.  

At least when all $R_i$ are small compared to the $11$ dimensional Planck
scale $l_p$, the low energy dynamics are described by an $11-k$ dimensional
supergravity theory with $32$ supercharges.
This theory is determined by the $k+1$ parameters $l_p, R_1, \dots, R_k$ which are positive real numbers, so naively one would expect
its moduli space to be
$\R_+^{k+1}$.  However, if we want the space of physically inequivalent theories then we have to take the quotient
of $\R_+^{k+1}$ by the duality group consisting of transformations which leave the physics invariant.  

For M-theory on a $T^k$ of arbitrary shape with arbitrary
$C$-field the duality group is well known to be
$E_{k(k)}(\Z)$ \cite{Hull:1995ys}.  In our case we want to
 consider only dualities which respect the
``rectangular, no $C$-field'' condition.
These include the symmetric group
$S_k$ permuting the $k$ radii; this group is generated by the $k-1$ exchanges
\begin{equation}
w_{\alpha_i}:  R_i \leftrightarrow R_{i+1}, \,\,i = 1, \dots, k-1.
\end{equation}
In addition one also has the action of T-duality
(M2/M5 exchange, after $T^3$ compactification) which
 when expressed in M-theory language involves
three radii, so we need to include one extra generator in case $k \ge 3$, namely
\begin{equation}
w_{\alpha_k}:  2 \pi R_1 \mapsto \frac{l_p^3}{2\pi R_2 2 \pi R_3},
2 \pi R_2 \mapsto \frac{l_p^3}{2\pi R_1 2\pi R_3}, 2\pi R_3 \mapsto
\frac{l_p^3}{2\pi R_1 2\pi R_2}, l_p^3 \mapsto \frac{l_p^6}{2\pi
R_1 2\pi R_2 2\pi R_3}.
\end{equation}
As we will see below, this corresponds to the quadratic transformation
discussed before
in the context of del Pezzos.
It is known \cite{Banks:1998vs, Obers:1998fb} that the $w_{\alpha_i}$ generate the full duality
group acting on ``rectangular, no $C$-field'' compactifications;
just as in Section \ref{weyl} this
group is isomorphic to the Weyl group of $E_k$.  Moreover, the other
elements of the U-duality group,
such as periodicity of the $C$ field or shifting
complex moduli of tori by $\tau \rightarrow \tau +1$, are manifest symmetries of
string theory.  So the non-obvious part of the
full U-duality group is already captured by the Weyl group $W(E_k)$.

Now we are ready to map the moduli.  Using
the notation $\mhk$ for the naive moduli space $\R^{k+1}_+$, we may write
\begin{equation}
\mk = \mhk / W(E_k).
\end{equation}
It is convenient to switch to a linear
representation by taking logarithms:  namely, we think of
$\mhk$ as a $(k+1)$-dimensional real vector space, with a typical element
\begin{equation}
(\log l_p, \log (2\pi R_1), \dots, \log (2\pi R_k)).
\end{equation}
Then $U_k$ acts linearly on $\mhk$.
To establish our correspondence we need one more piece of structure on $\mhk$.
Namely, by considering the
log-tension formulas for $\half$-BPS states
we obtain a lattice $\Lambda$ in the dual vector space
$\mhk^*$, spanned by $k+1$ basis vectors $(3 \log l_p)$
and $(\log 2\pi R_a)$ ($a = 1, \dots, k$).

In sum, we have a $(k+1)$-dimensional vector space $\mhk$,
 carrying an action of the Weyl group $U_k$ which
furthermore preserves a lattice $\Lambda$ in the dual space.
Precisely this structure is also present on the del Pezzo surface:  namely,
we have the $(k+1)$-dimensional cohomology $H^2(\B_k, \R)$, carrying
 an action of the Weyl group
by global diffeomorphisms preserving the canonical class, which
 furthermore preserves the homology lattice $H_2(\B_k, \Z)$.
The natural thing to do, then, is to identify the two vector
spaces $H_2(\B_k, \R)$ and $\mhk^*$ in a way which identifies
the lattices $H_2(\B_k, \Z)$ and $\Lambda$ while preserving the
 action of the Weyl group.
This can be done in an essentially unique way:
in the notation of Section \ref{homology}, we map
\begin{equation}
\begin{aligned} \label{homology-map}
H \mapsto & - 3 \log l_p, \\
E_a \mapsto & - \log (2\pi R_a).
\end{aligned}
\end{equation}
Dually, we have an identification between $\mhk$ and $H^2(\B_k, \R)$, described by
\begin{equation} \label{cohomology-map}
(l_p, R_1, \dots, R_k) \leftrightarrow (\omega \in H^2(\B_k, \R): \ \omega(H) = -3 \log l_p, \omega(E_a) = -\log (2\pi R_a)).
\end{equation}
We think of $\omega \in H^2(\B_k, \R)$ as a kind of generalized K\"ahler class.
If $\omega$ came from an ordinary (positive) K\"ahler metric, then
$\omega(\CC)$ would be simply the volume of a holomorphic
curve in the class $\CC$.  Our $\omega$ need not come from a K\"ahler metric,
since $\omega(\CC)$ may be negative (e.g. if $2\pi R_a>1$ then $\omega(E_a)<0$.)

The choice \eqref{homology-map} is not quite unique --- apart from the unimportant freedom
to make a Weyl group transformation on one side, we could also have
taken plus signs instead of minus.  Our choice was dictated by the original intuition
that blowing up $E_a$ corresponds to compactification from $R_a = \infty$ to $R_a = \mathrm{finite}$,
so that the volume of $E_a$ should be inversely related to $R_a$.

Before going further, let us remark on the appearance of the group of
diffeomorphisms
preserving the canonical class $K$ in the above correspondence.  Recall from \eqref{canonicalclass} that
\begin{equation}
K = - 3H + \sum_{a=1}^k E_a.
\end{equation}
Applying \eqref{homology-map} we find the correspondence
\begin{equation} \label{can-map}
K \mapsto (9-k) \log \pl,
\end{equation}
where $\pl$ is the Planck length in the compactified theory, which is
given by
\begin{equation}
\pl^{9-k} = \frac{l_p^9}{(2\pi R_1) \cdots (2\pi R_k)}.
\end{equation}
Since $\pl^{9-k}$ can be observed in the dimensionally reduced theory as the
inverse of the  gravitational
coupling constant, it must be invariant under every element in the U-duality group;
this is why we have to consider only diffeomorphisms
preserving $K$.

\subsection{Brane charges and rational curves}

Next we consider the tensions of $\half$-BPS states in M-theory.  We will
show that a holomorphic rational curve $\CC$ in the del Pezzo corresponds
to a $\half$-BPS $p$-brane charge.
Moreover,
the BPS tension can be identified as
\begin{equation} \label{tension}
T= 2 \pi \exp \omega(\CC),
\end{equation}
where $\omega \in H^2(\B_k, \R)$ is
the generalized K\"ahler class introduced in the last section.
Given this formula for the tension,
one easily checks that $p$ is determined by the intersection product with $K$
(recall that $K$ corresponds to $(9-k) \log \pl$, so we can think of it as in some sense
``carrying the units of mass''):
\begin{equation} \label{mass-dimension}
\degC=-\CC \cdot K=p+1.
\end{equation}
Furthermore, electric-magnetic duality has a simple interpretation in
this framework:  dual pairs are related simply by
\begin{equation} \label{elmag}
\CC_{\mathrm{Electric}} + \CC_{\mathrm{Magnetic}} = -K.
\end{equation}
In particular this will imply, using \eqref{tension} and \eqref{can-map}, that
\bea
T_{\mathrm{Electric}}T_{\mathrm{Magnetic}}=(2 \pi)^{2} \pl^{k-9},
\eea
for all electric-magnetic pairs of $\half$-BPS branes.

Before considering the general setup, let us consider the simple
case of uncompactified M-theory.

\medskip
\noindent {\bf{M-theory in $d=11$.}}
Recall that this theory corresponds to $\PP^2$.
The lattice $H_2(\PP^2, \Z)$ is spanned by the single element $H$, so
consider a curve in $\PP^2$ given by the class
\begin{equation}
\CC = n\,H.
\end{equation}
A priori we could choose any integral value for $n$.  However, not
every class in $H_2(\PP^2, \Z)$ is actually realized by a holomorphic
curve: one has to choose $n > 0$.  Choosing $n=1$, $H$ is the class of
a line in $\PP^2$, and the tension formula gives $T=2 \pi \exp
\omega(H)=2 \pi/l_p^{3}$.  Similarly, $2H$ is the class of a
conic and gives $T = 2 \pi/l_p^{6}$, the tension of the M5
brane.  These are the only genus zero curves in $\mathbb{P}^{2}$, since by \eqref{adjunction} 
the genus of $\CC=nH$ is given by
\bea
g(\CC)=\frac{(n-1)(n-2)}{2}\,.
\eea
The toric representations of $H$ and $2H$ are shown in \figref{m2m5}.
\onefigure{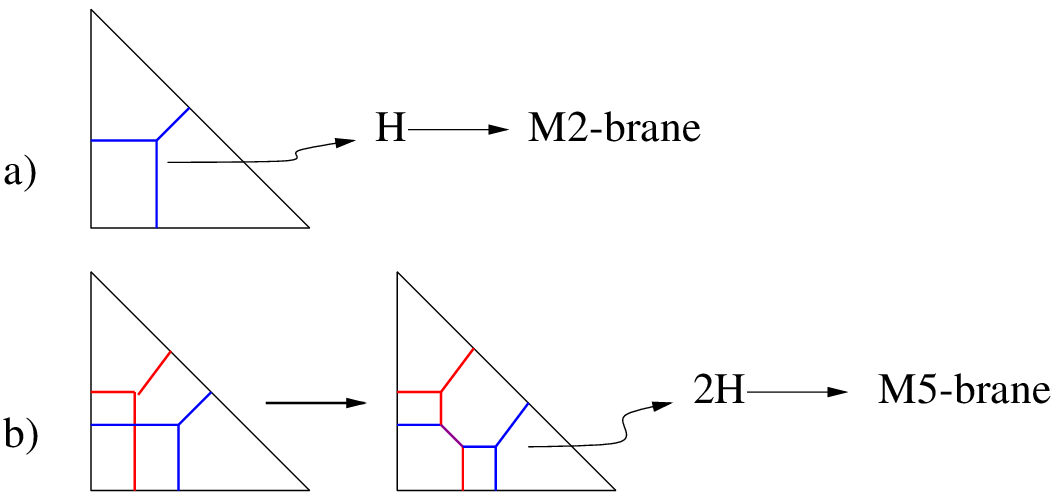}{a) Hyperplane class $H$ in $\mathbb{P}^{2}$ which
maps to the M2-brane, b) The conic $2H$ in $\mathbb{P}^{2}$ which maps to
the M5-brane.}  Note that the fact that the M2-brane and M5-brane are 
electric-magnetic duals agrees with the prediction from \eqref{elmag},
because $H+2H=3H=-K_{\PP^{2}}$.

What about $3H$?  This is the
class of a cubic curve in $\PP^2$.
If it plays a role in our correspondence it
should correspond to a tension $T = 2 \pi/l_p^{9}$, giving some hypothetical M8 brane.  On the other hand, as remarked earlier,
a cubic curve in $\PP^2$ has genus one.  So if we restrict to rational curves we obtain only the M2 and M5.

It is intriguing that the fact that the worldvolume dimensions
of the $\half$-BPS states in $11$ dimensions are a multiple of $3$ directly follows
from the fact that the canonical class of $\PP^2$ is
a multiple of $3$ given by $-3H$.
This is indeed a remarkable map!

\medskip

Before going on to a more complicated example, we discuss
some general aspects of the relevant physics.
States which are $\half$-BPS satisfy a formula
relating their tensions to the SUSY central charges.
These charges in turn are proportional to gauge charges
 in $11-k$ dimensions, with the constant of proportionality
depending on the parameters $R_a$ and $l_p$.  For example,
a state with one unit of momentum in the compact
$1$ direction would satisfy
\begin{equation} \label{mf-1}
M = \frac{1}{R_1},
\end{equation}
while an M2-brane wrapped
on the $1-2$ directions would have
\begin{equation} \label{mf-2}
M = \frac{(2 \pi)^{3} R_1 \,R_2}{l_p^3}.
\end{equation}
By ``gauge charge'' we include charges under $p$-form symmetries as
well as $1$-forms, so we can also consider e.g. an M5 wrapped on the $1-2-3$ directions.
In this case we get a tension formula rather than a mass formula, which reads
\begin{equation} \label{mf-3}
T = \frac{(2 \pi)^{4}\, R_1\, R_2\, R_3}{l_p^6}.
\end{equation}
All three tension formulas \eqref{mf-1}, \eqref{mf-2}, \eqref{mf-3}
are of the form
\begin{equation} \label{mf-general}
T = 2 \pi \frac{(2\pi R_1)^{m_1} \cdots (2\pi R_k)^{m_k}}{l_{p}^{3n}},
\end{equation}
with $n$ and all $m_{a}$ integral.
It is not quite true that this is the most general
 situation for a $\half$-BPS state --- for example, a state
with one unit of momentum in each of the $1$ and $2$ directions
would have $M = \sqrt{1 / R_1^2 + 1 / R_2^2}$.  Put differently,
the two states given by one unit of momentum in each of the $1$ and
$2$ directions form a bound state which has mass smaller than the sum
of the two masses.  We will discuss below the geometric condition
under which a pair of $\half$-BPS states form a bound state;
at this point we concentrate on a particular
basis for the lattice of charges, such that every state in the basis has a tension
formula of the form \eqref{mf-general}.

So let us relate the formula \eqref{mf-general} to the del Pezzo surface $\B_k$.  Setting
\begin{equation} \label{volume}
V = -3n \log l_p + \sum_{a=1}^{k} m_a \log (2\pi R_a),
\end{equation}
we can rewrite \eqref{mf-general} as
\begin{equation}
T = 2 \pi \exp V.
\end{equation}
On the other hand, through the map \eqref{cohomology-map} we can rewrite \eqref{volume} as
\begin{equation}
V = \omega(\CC),
\end{equation}
where
\begin{equation} \label{curve-form}
\CC = nH - \sum_{a=1}^k m_a E_a.
\end{equation}
So we have expressed the log-tensions of certain $\half$-BPS states as generalized volumes of particular classes in
$H_2(\B_k, \Z)$, using \eqref{homology-map} as our dictionary.

We can now ask the question:  which classes
actually do correspond to $\half$-BPS states?  We will see below that
the rational curves which correspond to $p$-branes with codimension
2 or more are in 1-1 correspondence with the $\half$-BPS brane charges
in M-theory.  We have already discussed the situation for M-theory
in 11 dimensions.  Let us now consider its compactification to 10 dimensions.

\medskip
{\noindent \bf{Type IIA in $d=10$.}}
Compactifying M-theory on a circle corresponds to blowing up a point on $\PP^2$ to get $\B_1$.
The homology lattice is then two-dimensional, so an arbitrary element
in $H_2(\B_1, \Z)$ is given by
\begin{equation}
\CC = nH - mE.
\end{equation}
Taking our cue from the result in case $k=0$, let us look for $\CC$ which are realized by holomorphic curves of genus zero.
The adjunction formula \eqref{adjunction} in this case becomes
\begin{equation}
n(n-3) + m(m-1) = 2g-2.
\end{equation}
Looking for integer solutions of the resulting quadratic,
we find two infinite families:
\begin{equation}
\begin{aligned}
(n,m) = \left(\frac{p}{2}, \frac{p}{2}-1\right)  & \ \ (p \in 2\Z) \\
(n,m) = \left(\frac{p+3}{4}, -\frac{p-5}{4}\right) & \ \ (p \in 4\Z+1).
\label{iia}
\end{aligned}
\end{equation}
Since we do not have branes of arbitrary dimension in type IIA,
we now impose a further restriction:  namely, we consider only
solutions with $0 \le p \le 8$.
Upon so doing we obtain the following list:

\vglue 0.5cm
\noindent
\begin{tabular}{||c|c|c||} \hline
homology class & tension & type IIA meaning \\ \hline \hline
$E$ & $R^{-1} = l_s^{-1} g_s^{-1}$ & D0-brane \\ \hline
$H-E$ & $(2\pi)^{2}\,R l_p^{-3} =(2\pi)^{-1} l_s^{-2}$ & F-string \\ \hline
$H$ & $(2\pi)\,l_p^{-3} = (2\pi)^{-2}\,l_s^{-3} g_s^{-1}$ & D2-brane \\ \hline
$2H-E$ & $(2\pi)^{2}\,R l_p^{-6} = (2\pi)^{-4}\,l_s^{-5} g_s^{-1}$ & D4-brane \\ \hline
$2H$ & $(2\pi)\,l_p^{-6} = (2\pi)^{-5}\,l_s^{-6} g_s^{-2}$ & NS5-brane \\ \hline
$3H-2E$ & $(2\pi)^{3}\,R^2 l_p^{-9} = (2\pi)^{-6}\,l_s^{-7} g_s^{-1}$ & D6-brane \\ \hline
$4H-3E$ & $(2\pi)^{4}\,R^3 l_p^{-12} = (2\pi)^{-8}\,l_s^{-9} g_s^{-1}$ & D8-brane \\ \hline
\end{tabular}

\vglue 0.5cm

Remarkably, we have all the relevant $\half$-BPS objects in type
IIA in 10 dimensions, with the correct tensions.  
To see this,
consider the $p$-brane which maps to the curve $\CC_{n,m}=nH-mE$, for some
$n$ and $m$ satisfying (\ref{iia}). The
tension is given by
\begin{equation} \label{above}
\begin{aligned}
T_{n,m}&=2\pi\,\mbox{exp}(\omega(\CC_{n,m}))\,,\\ &=2\pi
\,\frac{(2\pi R)^{m}}{l_{p}^{3n}}\,. 
\end{aligned}
\end{equation}
From \eqref{above}
it follows that\footnote{We use the relations
$R=g_{s}l_{s}$,\,\,$l_{p}=2\pi\,g_{s}^{\frac{1}{3}}l_{s}$
between M-theory parameters $l_p,R$ and type IIA parameters $l_s,g_s$ \cite{Schwarz:1998mm}.}
\begin{equation}
\begin{aligned}
T_{Dp-brane}&=T_{\frac{p}{2},\frac{p}{2}-1}=\frac{(2\pi)^{\frac{p}{2}}\,R^{\frac{p}{2}-1}}{l_{p}^{\frac{3p}{2}}}\,=\,\frac{1}{(2\pi)^{p}\,g_{s}\,l_{s}^{p+1}}\,,\\
T_{NS
p-brane}&=T_{\frac{p+3}{4},\frac{5-p}{4}}=\frac{(2\pi)^{\frac{9-p}{4}}\,
R^{\frac{5-p}{4}}}{l_{p}^{\frac{2p+9}{4}}}\,=\,\frac{1}{(2\pi)^{p}\,g_{s}^{\frac{p-1}{2}}\,l_{s}^{p+1}}\,.
\end{aligned}
\end{equation}
Thus we get correct tensions for all the D$p$-brane as well as
the fundamental string and the NS5-brane, as shown in the above table.

The electric/magnetic
pairing works as described by \eqref{elmag}, with 
\begin{equation}
\CC_{\mathrm{Electric}}+\CC_{\mathrm{Magnetic}}=-K=3H-E.  
\end{equation}
It is also interesting that the del Pezzo knows about
ALF space, i.e. about the D6 brane, which does not come from simple
dimensional reduction from 11 dimensions.  Similarly the appearance of the D8 brane here,
which does not correspond to any object in $\PP^2$, is remarkable.

\medskip
{\noindent \bf{Type II in $d=8$.}}
For brevity we now skip $d=9$ and go directly to $d=8$.  We search for curves which could represent
$p$-branes with $0 \le p \le 6$; using \eqref{adjunction} as above and \eqref{mass-dimension}, this amounts to finding integer solutions of
\begin{equation}
\begin{aligned}
3n - m_1 - m_2 - m_3 &= p + 1, \\
n^2 - m_1^2 - m_2^2 - m_3^2 &= p - 1.
\end{aligned}
\end{equation}
The results are summarized in the table below, where we
include the interpretation of the states
in terms of M-theory as well as Type IIA.  (One could also interpret these states from
the perspective of compactified Type IIB, using the relations \eqref{homology-identification}
between blow-ups of $\PP^2$ and of $\PP^1 \times \PP^1$.)

\vglue 0.5cm
\noindent
\begin{tabular}{|c|c|c|c|} \hline
$p$ & homology class & type IIA meaning & M-theory meaning \\ \hline \hline
\multirow{4}{*}{$p=0$} & $E_M$ & D0-brane & \multirow{2}{*}{momentum} \\ \cline{2-3}
& $E_i$ & momentum & \\ \cline{2-4}
& $H - E_i - E_j$ & twice-wrapped D2-brane & \multirow{2}{*}{twice-wrapped M2-brane} \\ \cline{2-3}
& $H - E_M - E_i$ & once-wrapped F-string & \\ \hline \hline
\multirow{2}{*}{$p=1$} & $H-E_M$ & F-string & \multirow{2}{*}{once-wrapped M2-brane} \\ \cline{2-3}
& $H-E_i$ & once-wrapped D2-brane & \\ \hline \hline
\multirow{2}{*}{$p=2$} & $H$ & D2-brane & M2-brane \\ \cline{2-4}
& $2H-E_M-E_i-E_j$ & twice-wrapped D4-brane & thrice-wrapped M5-brane \\ \hline \hline
\multirow{2}{*}{$p=3$} & $2H - E_M - E_i$ & once-wrapped D4-brane & \multirow{2}{*}{twice-wrapped M5-brane} \\ \cline{2-3}
& $2H - E_i - E_j$ & twice-wrapped NS5-brane & \\ \hline \hline
\multirow{4}{*}{$p=4$} & $2H - E_M$ & D4-brane & \multirow{2}{*}{once-wrapped M5-brane} \\ \cline{2-3}
& $2H - E_i$ & once-wrapped NS5-brane & \\ \cline{2-4}
& $3H - 2E_M - E_i - E_j$ & twice-wrapped D6-brane & \multirow{2}{*}{twice-wrapped ALF space} \\ \cline{2-3}
& $3H - E_M - 2E_i - E_j$ &  exotic state  & \\ \hline \hline
\multirow{5}{*}{$p=5$} & $2H$ & NS5-brane & M5-brane \\ \cline{2-4}
& $3H - 2E_M - E_i$ & once-wrapped D6-brane & \multirow{3}{*}{once-wrapped ALF space} \\ \cline{2-3}
& $3H - E_M - 2E_i$ & exotic state & \\ \cline{2-3}
& $3H - 2E_i - E_j$ & exotic state & \\ \cline{2-4}
& $4H - 2E_M - 2E_i - 2E_j$ & exotic state & exotic state \\ \hline \hline
\multirow{4}{*}{$p=6$} & $3H - 2E_M$ & D6-brane & \multirow{2}{*}{ALF space} \\ \cline{2-3}
& $3H - 2E_i$ & exotic state & \\ \cline{2-4}
& $4H - 3E_M - E_i - E_j$ & twice-wrapped D8-brane & \multirow{2}{*}{exotic state} \\ \cline{2-3}
& $4H - E_M - 3E_i - E_j$ & exotic state & \\ \hline
\end{tabular}
\vglue 0.5cm

In the above table we have singled out one of the exceptional curves as the
``M-theory direction'' and called it $E_M$, while the indices $i,j$ run over the values $1,2$ for
the other two circles.  The entries labeled ``exotic state'' correspond to states satisfying exotic
tension formulas which are not straightforward to interpret; nevertheless, these
states are required by the U-duality symmetry, as reviewed e.g. in \cite{Obers:1998fb}.

From looking at the table we notice a simple pattern, namely, if $\CC$ is a $\half$-BPS object not involving
$E_i$, then $\CC - E_i$ will be the same object wrapped on the $i$ direction.  One can also see this directly
from the tension formulas since $-E_i$ corresponds to $\log 2\pi R_i$.

\medskip
{\noindent \bf{Type IIB in $d=10$.}}
In our discussion so far we have mostly stuck to the surfaces $\B_k$, but the discussion goes
through essentially unchanged for $\PP^1 \times \PP^1$.  Recall that a basis for $H_2(\PP^1 \times \PP^1, \Z)$
is given by the classes of the two factors, which we write $l_1$ and $l_2$, with
\begin{equation}
\begin{aligned}
l_1 \cdot l_1 = l_2 \cdot l_2 &= 0, \\
l_1 \cdot l_2 &= 1.
\end{aligned}
\end{equation}
The canonical class is $K = -2l_1-2l_2$, and the analog of \eqref{homology-map} in this setting is
\begin{equation}
\begin{aligned}
l_1 &\mapsto -2 \log 2 \pi l_s, \\
l_1 &\mapsto -2 \log 2 \pi l_s - \log g_s,
\end{aligned}
\end{equation}
where $l_s$ and $g_s$ now denote the Type IIB quantities. 
From \eqref{defdeg} and \eqref{adjunction}, a genus zero curve ${\cal
C}=nl_{1}+ml_{2}$ of degree $p+1$ satisfies
\bea 2nm=p-1\,,\,\,\,2(n+m)=p+1\,. \eea There are two integer solutions for
each $p\in 2\Z+1$: 
\begin{equation}
\begin{aligned}
(n,m)&=\left(1,\frac{p-1}{2}\right)\,,\\
(n,m)&=\left(\frac{p-1}{2},1\right)\,. 
\end{aligned}
\end{equation}
 The tension is then given by
\begin{equation}
\begin{aligned}
T_{Dp-brane}&=T_{\frac{p-1}{2},1}=\frac{1}{(2\pi)^{\frac{p-1}{2}}}\,
T_{F-string}^{\frac{p-1}{2}}\,T_{D-string}=\frac{1}{(2\pi)^{p}\,g_{s}\,l_{p}^{p+1}}\,,\\
T_{NS p-brane}&=T_{1,\frac{p-1}{2}}=
\frac{1}{(2\pi)^\frac{p-1}{2}}\,T_{F-string}\,T_{D-string}^{\frac{p-1}{2}}=\frac{1}{(2\pi)^{p}\,g_{s}^{\frac{p-1}{2}}\,l_{p}^{p+1}}\,.
\end{aligned}
\end{equation}
Then restricting to $0 \le p \le 8$, we obtain

\vglue 0.5cm
\noindent
\begin{tabular}{||c|c|c||} \hline
homology class & tension / $2\pi$ & type IIB meaning \\ \hline \hline
\rule{0mm}{5mm}$l_1$ & $(2\pi l_s)^{-2}$ & F-string \\ \hline
\rule{0mm}{5mm}$l_2$ & $(2\pi l_s)^{-2} g_s^{-1} $ & D-string \\ \hline \hline
\rule{0mm}{5mm}$l_1+l_2$ & $(2\pi l_s)^{-4} g_s^{-1}$ & D3-brane \\ \hline \hline
\rule{0mm}{5mm}$2l_1+l_2$ & $(2\pi l_s)^{-6} g_s^{-1}$ & D5-brane \\ \hline
\rule{0mm}{5mm}$l_1+2l_2$ & $(2\pi l_s)^{-6} g_s^{-2}$ & NS5-brane \\ \hline\hline
\rule{0mm}{5mm}$3l_1+l_2$ & $(2\pi l_s)^{-8} g_s^{-1} $ & D7-brane \\ \hline
\rule{0mm}{5mm}$l_1+3l_2$ & $(2\pi l_s)^{-8} g_s^{-3}$ & NS7-brane \\ \hline
\end{tabular}
\vglue 0.5cm

Note that the fact that the worldvolume dimension of the Type IIB
$\half$-BPS branes are all even follows from the fact that $K$ for
$\PP^1 \times \PP^1$ is even, i.e.
$K=-2l_1-2l_2$.

\medskip
{\noindent \bf{Type II in general $d \ge 3$.}}
In parallel to the discussion above, one can check that all the
$\half$-BPS $p$-branes
with codimension at least 1 in uncompactified
spacetime in compactifications
of M-theory on $T^k$ for $k\leq 8$ are in 1-1 correspondence
with rational curves $\CC$
with $\degC = p+1$.  One finds all the branes which appeared in the
table for $d=8$, plus
their compactifications, plus various extra exotic states (again,
all required to exist by U-duality.) It is not actually difficult
to determine genus zero curves of a given degree satisfying the
adjunction formula. We can write the self-intersection of $\CC$
in terms of the weight vector $\CC_\perp$ as discussed in
section 2.2,
\bea
\CC^{2}=d_{\CC}-2\,\,\implies \CC_{\perp}^{2}=-\frac{d_{\CC}^{2}}{9-k}+d_{\CC}-2\,.
\label{condition}
\eea
The results are summarized in the following table in which we give the
representation of $E_{k}$ to which the curve belongs, the size of the Weyl
orbit and a representative curve. The uncompactified dimension 
of spacetime is $d=11-k$.

\vglue 0.5cm
\noindent
\begin{tabular}{||c|c|c|c|c|c|c||} \hline
$k$ & $E_{k}$ & $d_{\CC}$ & $\CC_{\perp}^{2}$ & rep of $E_{k}$& Weyl orbit size & curve\\ \hline \hline

\multirow{6}{*}{$k=4$}& \multirow{6}{*}{$SU(5)$}
& \rule{0mm}{5mm}1 & $-\frac{6}{5}$ & ${\bf 10}$ &10   &$E_{1}$  \\ \cline{3-7}
&&\rule{0mm}{5mm}2 & $-\frac{4}{5}$ & ${\bf 5}$  &5   &$H-E_{1}$\\ \cline{3-7}
&&\rule{0mm}{5mm}3 & $-\frac{4}{5}$ & ${\bf 5}$  &5  &$H$      \\ \cline{3-7}
&&\rule{0mm}{5mm}4 & $-\frac{6}{5}$ & ${\bf 10}$ &10   &$3H-2E_{1}-E_{2}-E_{3}-E_{4}$\\  \cline{3-7}
&&\rule{0mm}{5mm}5 & $-2$           & ${\bf 24}$ &20 &$2H-E_{1}$ \\ \cline{3-7}
&&\rule{0mm}{5mm}6 & $-\frac{16}{5}$& ${\bf 15}$ &5  &$2H$ \\ \cline{3-7}
\hline \hline

\multirow{5}{*}{$k=5$}& \multirow{5}{*}{$SO(10)$}
& \rule{0mm}{5mm}1 & $-\frac{5}{4}$ & ${\bf 16}$ & 16 &$E_{1}$\\ \cline{3-7}
&&\rule{0mm}{5mm}2 & $-1$ &${\bf 10}$ &10 &$H-E_{1}$\\ \cline{3-7}
&&\rule{0mm}{5mm}3 & $-\frac{5}{4}$ & ${\bf 16}$ & 16 & $3H-2E_{1}-\sum_{i=2}^{5}E_{i}$\\ \cline{3-7}
&&\rule{0mm}{5mm}4 & $-2$& ${\bf 45}$ &40& $2H-E_{1}-E_{2}$\\ \cline{3-7}
&&\rule{0mm}{5mm}5 & $-\frac{13}{4}$&  ${\bf 144}$ &80 &$2H-E_{1}$\\ \hline \hline

\multirow{4}{*}{$k=6$}& \multirow{4}{*}{$E_{6}$} 
&\rule{0mm}{5mm}1 & $-\frac{4}{3}$ & ${\bf 27}$ &27 &$E_{1}$ \\ \cline{3-7}
&&\rule{0mm}{5mm}2 & $-\frac{4}{3}$ & ${\bf 27}$ &27 &$3H-2E_{1}-\sum_{i=2}^{6}E_{i}$\\ \cline{3-7}
&&\rule{0mm}{5mm}3 & $-2$           & ${\bf 78}$ &72 &$H$\\ \cline{3-7}
&&\rule{0mm}{5mm}4 & $-\frac{10}{3}$& ${\bf 351}$&216&$2H-E_{1}-E_{2}$\\ \hline \hline

\multirow{3}{*}{$k=7$}& \multirow{3}{*}{$E_{7}$} 
&\rule{0mm}{5mm}1 & $-\frac{3}{2}$ & ${\bf 56}$ & 56& $E_{1}$ \\ \cline{3-7}
&&\rule{0mm}{5mm}2 & $-2$&${\bf 133}$ &126&$H-E_{1}$\\ \cline{3-7}
&&\rule{0mm}{5mm}3 & $-\frac{7}{2}$&${\bf 912}$ &576&$H$\\ \hline \hline

\multirow{2}{*}{$k=8$}& \multirow{2}{*}{$E_{8}$} 
&\rule{0mm}{5mm}1 & $-2$ & ${\bf 248}$ &240& $E_{1}$ \\ \cline{3-7}
&&\rule{0mm}{5mm}2 &$-4$ &${\bf 3875}$ &2160& $H-E_{1}$\\ \hline \hline

\end{tabular}
\vglue 0.5cm
\medskip

\subsection{Bound states of 0-branes}
Consider two 0-branes, represented by rational curves
$\CC_1,\CC_2$.  Then we may ask:  can they
form a BPS bound state?  For simplicity let us restrict
our attention to compactifications to $d \ge 5$ dimensions.
Then there are three different possibilities:
i) they form a bound state at threshold which is still $\half$-BPS,
ii) they form a bound state, not at threshold, which is $\half$-BPS,
iii) they form a bound state at threshold which is $\qtr$-BPS.
The first possibility is exemplified by a bound state of a pair of D0-branes.
The second is exemplified
by a bound state of a D0-brane with a 2-cycle wrapped D2-brane,
and the last occurs e.g. for the bound state of a D0-brane with a
4-cycle wrapped D4-brane.  It is not
too difficult to show that on $\B_6$, $\CC_1 \cdot \CC_2$ is the only Weyl
invariant of a pair of exceptional curves and can only be $-1, 0$ or $1$.
These correspond to the above three possibilities:
$${\rm \half-BPS\ at \ threshold} \leftrightarrow \CC_1\cdot \CC_2=-1\ \ (\mathrm{i.e.}\
\CC_1=\CC_2)$$
$${\rm \half-BPS\ not\ at\ threshold}\leftrightarrow \CC_1\cdot \CC_2=0$$
$${\rm \qtr-BPS \ at \ threshold} \leftrightarrow \CC_1\cdot \CC_2=1.$$
The mass of the bound state can be written in all
three cases as
$$M_{12}=2 \pi |e^{V_1}+i^{(\CC_1\cdot \CC_2)^2-1} e^{V_2}|.$$
Representatives of these three different cases can be chosen from the list
of possible 0-brane classes \eqref{zerob}:
\bea
\CC_1\cdot \CC_2&=&-1\,,\qquad \CC_1=E_a\,, \quad \CC_2=E_a\,,\\ \nonumber
\CC_1\cdot \CC_2&=&\,\,\,\,0\,,\qquad \CC_1=E_a\,, \quad \CC_2=E_b\,,\\ \nonumber
\CC_1\cdot \CC_2&=&\,\,\,\,1\,,\qquad \CC_1=E_M\,, \quad \CC_2=2H-E_M-E_1-E_2-E_3-E_4\,.
\eea
\subsection{Changes of scale}

In our discussion to this point we have implicitly assumed that some
energy scale $\mu$ has been fixed, with all quantities measured in
units of $\mu$.  Indeed, such a choice
is necessary in order to make sense of our main
formula \eqref{tension}, $T = 2 \pi \exp \omega(\CC)$, which requires that $T$ be
dimensionless.  To understand the significance of this choice let us
write the $\mu$ dependence explicitly: then for a $p$-brane state
corresponding to a curve $\CC$,
\begin{equation}
\log (T/2 \pi \mu^{p+1}) = \omega(\CC).
\end{equation}
Now consider changing scale from $\mu$ to $\mu'$; then we will have
\begin{equation} \label{dimensionchange}
\log(T/{2 \pi \mu'}^{p+1}) = \log(T/{2 \pi \mu}^{p+1}) + (p+1)
\log (\mu/\mu') = \omega(\CC) + (p+1) \log (\mu/\mu').
\end{equation}
On the other hand, using \eqref{mass-dimension} we can also rewrite
\eqref{dimensionchange} as
\begin{equation}
\log(T/{2 \pi \mu'}^{p+1}) = \omega'(\CC),
\end{equation}
where
\begin{equation}
\omega'(\CC) = \omega(\CC) + \log(\mu/\mu') \CC \cdot K.
\end{equation}
Hence a change in $\mu$ is equivalent to a shift in the generalized K\"ahler class $\omega$ along the
direction of $K$.  In particular, shifts along the $K$ direction do not affect any dimensionless quantities
one might compute on the M-theory side (e.g. ratios of masses of $\half$-BPS states.)  
So it is natural to think of $\omega$ as
decomposed into
\begin{equation} \label{decomposition}
\omega = \omega_\perp + \lambda K
\end{equation}
where $\omega_\perp$ is orthogonal to $K$; then $\omega_\perp$ controls all the dimensionless quantities
and $\lambda$ sets the units of measurement.  Setting $\mu = \pl$ would correspond to fixing $\lambda = 0$ in
\eqref{decomposition}.

The situation just described is somewhat counterintuitive:  one might have expected that
the choice of units in M-theory would correspond to the overall volume of the del Pezzo, but the
exponential relation between volumes and masses spoils that idea.  In particular, given any
$\omega \in H^2(X, \R)$, there is some $\lambda$ for which $\omega + \lambda K$ is in the
K\"ahler cone; so if we interpret $\omega$ as a generalized metric,
then whether volumes of curves on the del Pezzo are positive or not is dependent on the units
we choose.

\section{Concluding discussion}

The duality we have described is mathematically rather striking.
Of course, it is important to uncover its {\it physical} interpretation.
It is hard to believe that the correspondence
is purely accidental (it is reminiscent of the purely ``accidental''
appearance of Dynkin diagrams in the intersection matrix
of vanishing 2-cycles in $K3$).  One possibility is that the del Pezzo
is the moduli space of some probe in M-theory.  If so this probe must
be a U-duality invariant probe, and should be unlike any brane with which
we are familiar.  However, we can obtain a hint about what it should be
from the del Pezzo side.  Recall that the U-duality group was mapped to the group of
global diffeomorphisms of del Pezzo which preserve $K$.  Thus
a curve in the class given by $-K$ will be U-duality
invariant. Such a curve would have genus one (it is the elliptic curve
in $\PP^2$, with the blow-up classes added.)
Up to now we have only considered rational curves, but
if we assume there is a $p$-brane associated to this duality invariant
elliptic curve, then we get
$$p+1=(-K)\cdot (-K)=9-k$$
i.e., it would be an $8-k$-brane, which would have codimension
2 in the uncompactified spacetime.  It would be interesting
to see if such a U-duality invariant brane exists in some sense,
and if so
whether its moduli space is given by the del Pezzo.

A rather interesting suggestion has been made by Motl:
perhaps the del Pezzos should be viewed as target spaces
of $(2,1)$ strings, thus providing a concrete
realization of the proposal 
\cite{Kutasov:1996fp}.
There are a number of obstacles to overcome to make
this precise, but there are encouraging signs
that this idea may be on the right track \cite{prog}.

Up to now, we have considered only rectangular compactifications
of M-theory, with the $C$ fields vanishing.  It is natural to ask how we can
relax these conditions.  If we fix $\pl$ then by a naive dimension count
we find that the tangent space to the moduli space of M-theory
compactifications could correspond to
$$\bigoplus_{p=1,2,3,6} \Lambda^p (H^2_\perp(\B_k)),$$
where the symbol
$\perp$ means we restrict to the subspace orthogonal to $K$.  The rectangular
compactifications correspond to $p=1$ above; this is the map
we have already described.  The $p=2$ can be
viewed as the choice of making parallelogram compactification of tori,
$p=3$ is turning on the $C$-field and $p=6$ corresponds to turning
on the 6-form field coupling to the M5-brane (for compactifications
to 3-dimensions some other exotic moduli appear).  It would be interesting
to interpret the $p\neq 1$ deformations in some geometric way.  Similarly,
it would be useful to have a clearer geometric understanding of the role
of $\half$-BPS states whose masses are not of the simple form \eqref{mf-general}.

It would also be interesting to extend this kind of duality to other
compactifications of M-theory.  A natural candidate to study
in this case is the duality web with 16 supercharges.

\section*{Acknowledgements}
We would like to thank Joe Harris, Deepee Khosla, John Morgan 
and Lubo\v{s} Motl for
valuable discussions.  AI would also like to thank Matthew J.
Strassler and Barton Zwiebach for valuable discussions and the
High Energy Theory Group at University of Pennsylvania for
hospitality where part of this work was done.  The research of AI
is supported by NSF grant PHY-0071512. The research of AN is
supported by an NDSEG Graduate Fellowship.  The research of CV is
supported in part by NSF grants PHY-9802709 and DMS-0074329.

\section*{Appendix: Group-theoretic description}

To emphasize the mathematical naturality of our construction,
we recall a more abstract description of the moduli space of M-theory
on $T^k$ with $\pl$ held fixed:  namely, the moduli space is of the form
$G(\Z) \backslash G / K$, where $G = E_{k(k)}(\R)$ is the maximally noncompact real form of
$E_k$, and $K$ is a maximal compact subgroup.  Restricting to rectangular tori with no $C$ field
corresponds to restricting to the Cartan subgroup $H \subset G$; the part of $G(\Z)$ that
preserves $H$ is isomorphic to the Weyl group $W$, and the quotient by $K$ restricts us to
the identity component $H_0$ of $H$, so abstractly we have
\begin{equation}
\M_k \iso H_0 / W.
\end{equation}

For example, in case $k=4$
we have $G = SL(5, \R)$.  Then $H$ consists of diagonal matrices with determinant
$1$, and $H_0$ consists of such matrices with diagonal entries positive.
The Weyl group is the symmetric group $S_5$, which acts by permuting the
diagonal entries; this is the manifestation of U-duality in this picture.

Now what is the meaning of the logarithms and exponentials which appeared in Section 3?
Given $H_0$ we can consider its (abelian) Lie algebra $\h$, and the exponential map
$\exp: \h \to H_0$.  Since we are working with the maximally non-compact form this map is in fact
a bijection, so it identifies $\h$ with $H_0$; furthermore it commutes with the action of $W$,
which acts linearly on $\h$.

Hence it is mathematically very natural to take logarithms to identify $\M_k$ with the linear space $\h$.  On the other hand
$\h$ contains a natural lattice, namely, the lattice of ``coweights,'' dual to the weight lattice in $\h^*$.
Furthermore $\h$ carries an orthogonal action of the Weyl group.
This is essentially the structure which was uncovered in
Section \ref{moduli-spaces} and identified with $H_2(\B_k, \Z) \subset H_2(\B_k, \R)$
(except that in that section we had a lattice of 
dimension $k+1$ rather than $k$; the one-dimensional extension
arises because we allow $\pl$ to vary in order to sweep out all of $H_2(\B_k, \R)$.)
The fact that the tension formulas involve the exponential is also natural from this point of view
since the BPS mass formulas
can be expressed in terms of the eigenvalues of the $E_k$ action on states
\cite{Witten:1995ex}.

\bibliography{physics}

\end{document}